\pdfoutput=1
\documentclass[useAMS,,usenatbib]{mn2e}
\usepackage[usenames,dvipsnames]{color}
\usepackage{graphicx,bm,mathabx,braket,dsfont,hyperref}

\title[]{On the Evidence for Cosmic Variation of the Fine Structure
  Constant (I): A Parametric Bayesian Model Selection Analysis of the Quasar Dataset}
\author[E. Cameron and A. N. Pettitt]{E. Cameron$^{1}$\thanks{E-mail:
dr.ewan.cameron@gmail.com} and A. N. Pettitt$^{1}$\\
$^{1}$School of Mathematical
Sciences (Statistical Science), Queensland University of Technology
(QUT), GPO Box 2434,\\ Brisbane 4001, QLD, Australia}
\begin{document}

\date{Submitted to MNRAS: xx July 2013.}

\pagerange{\pageref{firstpage}--\pageref{lastpage}} \pubyear{2013}

\maketitle

\label{firstpage}

\begin{abstract}
We review the evidence behind recent claims of spatial variation in
the fine structure constant deriving from  observations of ionic absorption lines in the light
from distant quasars.  To this end we expand upon previous
non-Bayesian analyses limited by the assumptions of an unbiased and
strictly Normal distribution for the  ``unexplained errors'' of the benchmark quasar
dataset.  Through the technique of reverse logistic regression we estimate and compare marginal likelihoods for three
competing hypotheses---\textsc{(i)} the null hypothesis (no cosmic
variation), \textsc{(ii)} the monopole hypothesis (a constant
Earth-to-quasar offset), and \textsc{(iii)} the monopole+dipole
hypothesis (a cosmic variation manifest to the Earth-bound observer as
a North--South divergence)---under a variety of candidate parametric forms for the
unexplained error term.  Our analysis reveals weak support for a
skeptical interpretation in which the
apparent dipole effect is driven solely by systematic errors of opposing sign
inherent in measurements from the two telescopes employed to obtain
these observations.  Throughout we seek to exemplify a `best
practice' approach to Bayesian model selection with prior-sensitivity
analysis; in a companion paper we extend
this methodology to a semi-parametric framework using the infinite-dimensional Dirichlet
process.
\end{abstract}


\section{Introduction}\label{introduction}
Recent claims by  \citet{web11} and \citet{kin12} of a cosmic dipole signal in the fine structure
constant deriving from their extensive compilation of Keck and VLT quasar absorption
line measurements have been
greeted with a healthy mix of excitement and
skepticism by cosmologists at large. While undoubtably controversial with respect to the
prevailing picture of a Copernican Universe, homogeneous
in its composition and physical laws on the largest scales, the possibility of
\textit{some} late-time variation in the ``fundamental constants'' has been
expressly identified within a number of well-studied cosmological
theories.  The
touchstone for skeptical reaction to these claims was in fact the
close alignment between the equator of the alleged dipole and the
North--South divide between the sightlines of the two
telescopes used to collect these data; the skeptical interpretation
being that systematic errors of opposing sign are to blame for the
apparent signal.
Crucially, while the presence of an unexplained 
error term in the quasar dataset has been acknowledged by the Webb et
al.\ team, to-date all estimates of statistical significance for the
dipole (i.e., those by the Webb et al.\ team and their
colleagues at UNSW; \citealt{ber12}) have been computed
under the assumption of strictly unbiased, Normal errors.
Hence, the motivation for our Bayesian re-analysis of the dataset
under a variety of biased and unbiased, Normal and non-Normal error models.

The Bayesian model selection (BSM) framework used herein aims to
identify the most plausible model to explain the observed data (with
the hope of minimizing the posterior predictive error) from
amongst a pre-defined set of hypotheses \citep{ber96,kad04}.  The
quantitative basis for the BSM procedure is, characteristically, a
ratio of marginal likelihoods \citep{jef61,jay03}; the resulting ``Bayes
factor''  operating much like an automatic
Ockham's Razor favouring simplicity over complexity (cf.\
\citealt{jef92}).  As a well-motivated, ``automatic'' procedure to distinguish between
rival theories, given even limited or heterogeneous data, BSM has become
highly popular in cosmological (and astronomical) research with novel applications
\citep{tro07,van09} now
abounding in the literature and user-friendly software packages for
marginal likelihood estimation \citep{wei12x,fer13} readily available.
However, with this powerful machinary at hand it can be all too easy to fall into the trap of na\"ively/lazily
applying the BSM technique without due regard for its limitations,
most notably the sensitivity of Bayes factors to the chosen priors on
the internal parameters of each candidate model.  Hence, in the
present study we give particular attention to demonstrating the key
elements of principled BSM with prior-sensitivity analysis \citep{kas95,gel03}; in this endeavour we
hope to provide a minimal template for future cosmological BSM
studies.  Though we restrict the present
investigation to the usual case of parametric model selection, 
in a companion work (Paper II) we extend our methodology to a potentially more challenging semi-parametric formulation.

The structure of this paper is as follows.  In the remainder of the
Introduction we give a detailed, historical overview of the observational evidence
for and against a cosmic variation in the fine structure constant.  In
Section \ref{quasardataset} we describe the Webb et al.\ team's
publically-available ``quasar dataset'' and give a preliminary
investigation into the nature of the unexplained error term.  In
Section \ref{generativemodels} we propose a number of candidate
mathematical forms for the latter, explain our choice of prior densities
on their governing parameters, and examine the resulting posterior
distribution of each.  In Section \ref{rlr} we briefly review the reverse
logistic regression procedure for marginal
likelihood estimation, before proceeding to
report and compare the latter for each hypothesis plus error model
pairing (and to conduct our prior-sensitivity analysis) in Section \ref{marginallikelihoods}.
Finally, we summarize our conclusions and discuss the merits of
possible follow-up observational strategies for resolving this debate in Section \ref{conclusions}. 

\subsection{The Fine Structure Constant in a Cosmological Context} Introduced in 1916 by Arnold Sommerfeld to abbreviate a
recurring, dimensionless factor in his relativistic extension of the Bohr model for the
hydrogen atom, the fine structure constant, $\alpha$, is now
recognized as one of the fundamental coupling terms of quantum
electrodynamics (QED).  In this context $\alpha$ serves to characterize the strength of the
electromagnetic interaction and has been humorously dubbed, ``the Peter
Sellers of QED'' \citep{bor86}, owing to the many different roles it
plays within this theory: setting the scale for all electromagnetic
cross-sections, binding energies, and decay rates; and, of course, the
scale of fine structure splitting in atomic and ionic spectral
lines (cf.\ \citealt{gri05}).

Defined by the ratio of the
squared elementary charge, $e^2$, to the permittivity of free space,
$\varepsilon_0$, the reduced Planck constant, $\hbar$, and the vacuum
speed of light, $c$,
\begin{equation}
\alpha = \frac{e^2}{(4 \pi \varepsilon_0)\hbar c},
\end{equation}
the fine structure constant has an on-Earth laboratory value, known to
remarkable precision through exacting experimentation on quantum scale
systems, of $\alpha^{-1}
\approx 137.035999037(91)$ \citep{bou11,han08}.  Though nominally
designated a ``constant'' there are various theoretical foundations to
support a time-/space-varying $\alpha$, if empirical evidence of
such can be definitively established; most notably, within electrodynamic scalar field models
\citep{bek82,car98} and grand unification theories \citep{mar84,bra03}.  An historical
evolution (with respect to cosmological time) of the fine structure constant driven by a decreasing speed of light could even
offer a compelling solution to the infamous ``Horizon
  Problem''\footnote{Namely, the remarkable homogeneity of the Universe
beyond even the scale of casual connection under the standard model; that is, over physical
separations exceeding the maximum distance traversable
since the beginning of time at the speed of light.} of Big
Bang cosmology without inflation \citep{mof93,bar99,alb99}.

At present there exist just a
handful of different approaches, both
terrestrial and astronomical, through which one may
search for this experimentalist's ``Holy Grail''.  These include: (\textsc{i}) the
in-laboratory comparison of optical atomic clocks \citep{for07,ros08}; (\textsc{ii})
 the nuclear modelling of samarium isotopes
extracted from the Oklo natural fission reactor
\citep{shl76,dam96,gou06} or rhenium extracted from Earth-fallen meteorite samples
\citep{oli04}; (\textsc{iii}) the  cosmological
modelling of angular fluctuations in the temperature and polarization
of the Cosmic Microwave Background \citep{roc04,nak08,men09,gal10,cal11}; and/or
(\textsc{iv}) the
identification of telltale frequency shifts in $\alpha$-sensitive features of
astronomical spectra, most notably the fine structure
emission lines of ionized carbon observed in radio waves from distant lensed
galaxies \citep{lev12,wei12} and the ionic absorption lines of various
species observed in
the visible/near-visible light from distant quasars
\citep{web99,web01,mur01,mur03,web11,kin12}.  After a number of initial disagreements between rival
teams---see \citet{lam04}, for instance---now resolved, only the
post-Millennial quasar-based studies at present claim to have
recovered any significant evidence for an evolving fine
structure constant.

First identified in 1966 \citep{bur66,sto66} the distinctive
absorption lines apparent to Earth-bound
observers at UV-to-optical wavelengths in the light from distant
quasars  arise from the
interaction of this light with ions of various species encountered during its passage through
intervening gas clouds; these residing in the
extensive halos of both
bright, star-forming galaxies and dark
proto-galaxies.  The potential of these absorption lines (especially
the singly and triply ionized silicon [Si\textsc{ii} and
Si\textsc{iv}] doublets) as probes of
the fundamental constants at extra-galactic distances,
complementary to the already-used
emission line technique \citep{sav56,bah65}, was
quickly realized by \citet{bah67} and used to
constrain the cosmic variation of $\alpha$ to within $\sim$10\% of its on-Earth value ($\Delta \alpha /
\alpha = 0.98$[0.05])\footnote{Throughout both the astronomical literature and our analysis
herein the particular notation, $\Delta
\alpha / \alpha$, is used to represent the fractional offset of the fine
structure constant in the extra-galactic system under study from its
on-Earth laboratory value, i.e.,
$\Delta
\alpha / \alpha = [\alpha - \alpha_\mathrm{Earth}]/\alpha_\mathrm{Earth}$.}  in the direction of one particular quasar (3C
191). Improvements in astronomical
instrumentation have since allowed far stronger
constraints on $\alpha$ variation to be established through this
approach. \citet{iva99}, for instance, have demonstrated $|\Delta \alpha/\alpha| < 2.3 \times
10^{-4}$ (95\% CI) in the early Universe (at $\sim$4 Gyr after the Big
Bang; $\sim$9.4 Gyr ago) through Si\textsc{iv} doublet observations along nine quasar
sightlines from a ground-based Russian telescope (the BTA-6
 at the Special Astrophysical Observatory).

To search for cosmic $\alpha$ variation below this limit (of roughly one
part in ten thousand) with quasar absorption line spectra requires
application of the ``many multiplet'' (MM) technique \citep{dzu99} in which the
relative frequency shifts of \textit{multiple} ionic species are
simultaneously compared; the transitions of those species
relatively insensitive to
 $\alpha$ variation (e.g.\ singly ionized magnesium [Mg\textsc{ii}]) effectively serving as
 calibration benchmarks for the stronger shifters (e.g.\ singly
 ionized iron [Fe\textsc{ii}]). Implementation of the MM technique necessarily
involves a multi-parameter fit to constrain not only $\Delta
\alpha / \alpha$ but also
a suite of
 nuisance parameters accounting for the physical structure of the
 intervening gas cloud (i.e., column
 density, kinetic temperature, and the dominant line broadening mechanism).
   Pioneering this technique in 1999 the Webb et al.\ team  \citep{web99} recovered tentative
evidence that $\alpha$ may have been lower in the past from a sample of 30 extra-galactic
absorbers spanning $0.5 < z_\mathrm{abs} < 1.6$ (here $z_\mathrm{abs}$
denotes the cosmological redshift of the absorber(s) under study)
observed with the Keck telescope ($\Delta \alpha
/\alpha = -1.1\ [0.4] \times 10^{-5}$).  This
exciting result was quickly heralded as support for a certain class of cosmological
model admitting a
decelerating speed of light \citep{bar00,dav02}, though the claimed theoretical
basis for
the well-publicized Davies et al.\ interpretation---derived from a
(mistaken) consideration of black hole thermodynamics---was soon disproven \citep{car03}.
More interestingly from a statistical point of view was the
Webb et al.\ team's concern for a possible underestimation of the true
errors in their quoted  
uncertainties, with a remarkably strong dip in their
$\alpha$ estimates over a narrow redshift interval ($0.9 \lesssim
z_\mathrm{abs} \lesssim 1.1$) suggesting the presence of an unexplained source of
observational error.

Further post-Millennial studies by the same team
\citep{web01,mur01,mur03} with the Keck telescope---expanding their original sample to a total of
141 absorbers and their redshift baseline to $0.2 < z_\mathrm{abs} < 3.7$---ultimately
strengthened the apparent weight of evidence  for a time-varying fine structure
constant beyond the ``4$\sigma$ level'';  given the particular assumptions
of the statistical analysis employed.  Chief amongst these the
absence of any \textit{bias} in the afore-mentioned unexplained error term---its
presence still inferred from a marked excess of the sample variance
over that expected from known sources of observational
noise, seemingly greatest within the high redshift
population.
At this time \citet{mur03} proposed a number of potential explanatory
factors for an additional error term unique
  to high redshift systems---including the entry of damped
  Lyman-alpha absorbers (dense clouds of neutral hydrogen featuring complex
  velocity structures more challenging to model via the MM technique) into the
  sample at $z_\mathrm{abs} \gtrsim 2$ (where the rest-frame
  ultra-violet 
  of their characteristic spectral lines is redshifted within the
  optical window accessible to on-Earth observers).  An alternative hypothesis, the
imprint of a spatial variation in $\alpha$ (as we describe below), was discounted at this stage as only weakly supported by
the available data (according to a bootstrap significance test); see \citet{mur03}.

\begin{figure*}
\vspace{0pc}
\begin{center}
\includegraphics[width=0.4\textwidth]{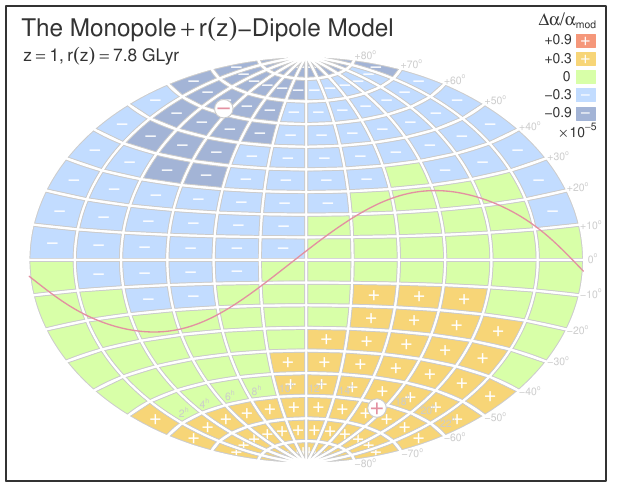} \includegraphics[width=0.4\textwidth]{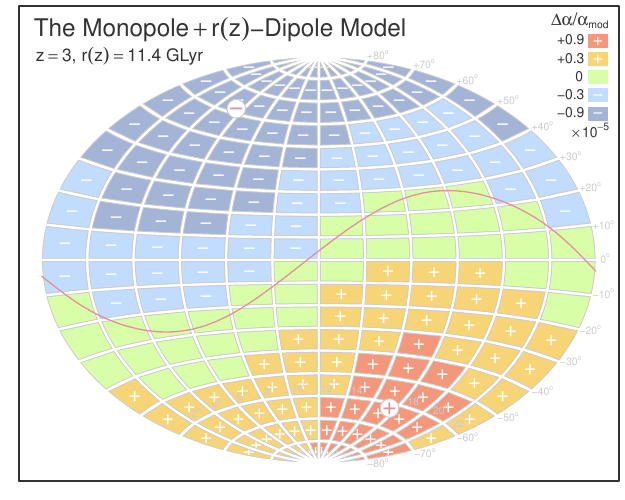}\end{center}
\vspace{-0.275cm}\caption{Visualization of the 
  monopole+$r(z)$-dipole model (cf.\ Section \ref{generativemodels}) for explaining the apparent spatial
  variation of the fine structure constant 
  proposed by the Webb et al.\ team.  The fractional difference of the fine
  structure constant from its on-Earth value under this model, $\Delta
  \alpha / \alpha_\mathrm{mod}$, as a function of the observational sightline for the
  best-fit solution of \citet{kin12} is shown at $z=1$  in the lefthand
  panel and that at $z=3$ in the righthand panel (with the
  color-/symbol-coding explained in the top-right legend of each).  The
  maximum, minimum, and equator of the key North--South dipole component of
  this model are
    overlaid as well for reference.}
\label{kingmodel}
\end{figure*}

It was thus a great surprise in 2010 when new $\Delta \alpha /
\alpha$ estimates for 154 (primarily Southern hemisphere) absorption
systems along 60 quasar sightlines (52 new and 8 in common with the original
Keck sample) derived by the same team \citep{kin12,web11}---but this
time using archival spectra from the Very Large Telescope (VLT) in Chile---appeared to indicate the opposite
evolutionary trend with redshift.
Namely, that the fine structure constant
was in fact higher in the past for these absorbers ($\Delta \alpha
/\alpha = 0.154\ [0.132] \times 10^{-5}$). To resolve this contradiction the team were forced to resurrect the spatial variation hypothesis,
proposing a smooth transition in $\alpha$ across the Universe
manifest to the on-Earth observer as a ($z$-invariant) monopole
plus a ($z$-dependent\footnote{\citet{kin12} in fact consider a
  variety of candidate functional forms for their dipole model,
  including a (redshift) $z$-invariant dipole, a $z^\beta$-dependent
  dipole, and an $r(z)$-dependent dipole (with $r(z)$ the cosmological
  lookback  distance).  All exhibit (bootstrap randomization-based)
  statistical significances of $\sim$4$\sigma$ over a monopole-only
null.  However, we note that: \textsc{(i)} a $z$-invariant dipole
 implies a strong breaking of the Copernican principle---namely, that
the Earth-bound observer does not occupy a privileged position within
the cosmos; and
\textsc{(ii)} the $z^\beta$-
alternative
(which encompasses a close approximation to the $r(z)$- model at
$\beta\approx0.3$) already appears from the \citet{kin12} study to be
an over-fitting of the available data.  Hence, we
focus exclusively on the (monopole+)$r(z)$-dipole scenario in the present
analysis.  For reference, this was also the approach taken by
\citet{ber12}.}) dipole field.  We
illustrate graphically in Figure \ref{kingmodel} the nature of the
spatial variation in $\Delta \alpha / \alpha$
under the best-fit model of this form (cf.\ Section \ref{generativemodels})
from \citet{kin12} on a color-/symbol-coded
map of the
celestial sphere.  Note the scale of the inferred variation,
which is at the $\Delta\alpha/\alpha \lesssim 10^{-5}$ level only
accessible (as noted earlier) via the MM technique.

\begin{figure*}
\vspace{0pc}
\includegraphics[width=0.825\textwidth]{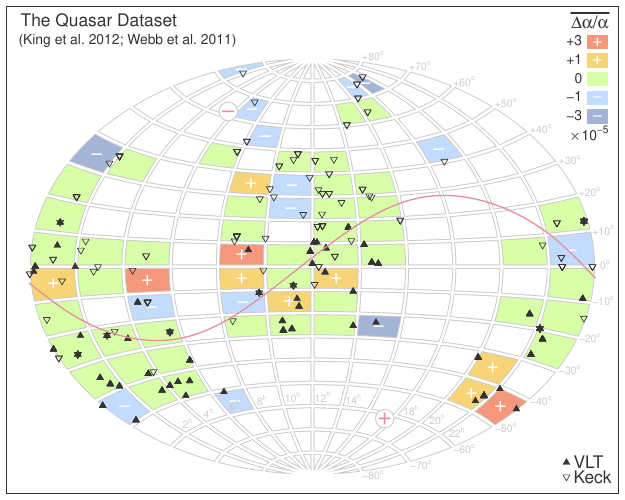}
\vspace{+0.1cm}\caption{Visualization of the apparent dipole signature in the
  \citet{kin12} / \citet{web11} quasar dataset.  Each quasar sightline
  in the sample is marked here on a projection of the sky based on the J2000 equatorial coordinate system; with solid-symbols
   denoting VLT observations and open symbols Keck observations.
   Note that for most of these quasars there are multiple intervening absorbers providing independent
   measurements of $\Delta \alpha / \alpha$ along the sightline.
   The apparent spatial variation of this observable is illustrated
    via a color-/symbol-coding (detailed in the top-right legend) of
    its weighted mean in each of the indicated
    subdivisions.  The maximum, minimum, and equator corresponding to
    the best-fit monopole+$r(z)$-dipole solution of the \citet{kin12} paper are
    overlaid as well for reference.}
\label{kingdipole}
\end{figure*}

In
Figure \ref{kingdipole} we present a comparable visualization of the
``raw'' $\Delta\alpha/\alpha$ variation in the Webb et al.\ team's
quasar dataset from which the apparent dipole
effect  was inferred.  To this end we
compute in each bin of right ascension and declination containing at least one quasar sightline (but typically two or
more; noting as well that there are on
average 2-3 absorbers per sightline) the weighted mean, \begin{equation} \overline{\Delta \alpha /
  \alpha} = \frac{\sum  (\Delta \alpha / \alpha_i) /
  (\sigma_{\mathrm{obs},i}^2+\sigma_{\mathrm{sys},i}^2)}{
\sum 1/(\sigma_{\mathrm{obs},i}^2+\sigma_{\mathrm{sys},i}^2)}.\end{equation}
Here  $\sigma_\mathrm{obs}$ and $\sigma_\mathrm{sys}$ represent the
standard deviations of the explained and 
unexplained error terms, respectively, in the Webb et al.\ team's
proposed generative model (which we detail in Section
\ref{quasardataset}).  Despite the substantial degree of noise evident
in these measurements (note the increase in scale with respect to that of Figure \ref{kingmodel})
one may yet discern an excess of negative $\Delta \alpha /
\alpha$ estimates in the far North and an excess of positive $\Delta \alpha /
\alpha$ estimates in the far South as per the dipole hypothesis.  Though, in anticipation of the
skeptical interpretation of these results (i.e., that biases of
opposite sign in the measurements from each telescope are to blame for
the apparent North--South divergence), we note also that the equator of the alleged dipole provides
a near perfect
subdivision of the sample into its VLT and Keck constituents.

A  ``4$\sigma$ level'' significance for the spatial variation hypothesis was calculated by \citet{kin12} through a bootstrap randomization of
$\Delta \alpha / \alpha$ estimates across sightlines and surveys
against the simple monopole (i.e., time-/space-invariant with
constant [non-zero] Earth-to-quasar offset) alternative---a result
echoed by \citet{ber12} in their review of the dataset using the
Akaike Information Criterion, the $F$ statistic, and the ``error
ellipsoid method''.  With the new VLT sample apparently just as
subject to unexplained measurement noise as the original Keck dataset,
however, a number of strong assumptions were required to justify the
significance testing procedures employed. In particular, it was
considered necessary to treat the aforesaid as
strictly Normal and strictly unbiased (viz.\ zero mean and
mode).  Assessing the impact of
these assumptions, which
are easily relaxed within a Bayesian framework as we demonstrate herein, thus represents an essential
``next step'' in the analysis of this observational benchmark given the profound
implications for both fundamental physics and cosmology if the dipole
interpretation is to become accepted.

Following the extensive world-wide media coverage of their work\footnote{See, for example, the 23 October 2010 issue of New
  Scientist magazine (\#2783), or the 2 September 2010 issue of The
  Economist magazine.} the Webb et al.\ team's spatial variation hypothesis was strongly criticized in a number of public forums, including the popular science
blogs, ``Uncertain Principles'' by \href{<http://scienceblogs.com/principles/2010/09/>}{Chad Orzel}
and ``Cosmic Variance'' by
\href{<http://blogs.discovermagazine.com/cosmicvariance/2010/10/18/the-fine-structure-constant-is-probably-constant/>}{Sean
Carroll}.
The
former citing the evident alignment of the alleged dipole's equator with
the coverage overlap of the Keck and VLT telescopes (cf.\ Figure \ref{kingdipole}) as an indicator that
biases of opposite sign in the observations from each
might well be to blame; and the latter citing the extraordinarily low
mass-energy budget ($\sim$$10^{-42}$ GeV) for the underlying scalar field implied by
the cosmic expanse of the fitted dipole.  From a Bayesian perspective these
objections may be viewed as disagreements over the relative prior
probabilities of competing proposals.  Namely, the degree to which the
particular model for the unexplained error term adopted by the
experimenters should be favoured over some alternative model (or
family of models), and the degree
to which the null hypothesis (of a time-/space-invariant $\alpha$)
should be favoured ``theoretically'' over the proposed dipole
hypothesis.

\section{The Quasar Dataset}\label{quasardataset}
In this study we examine the
publicly-available quasar dataset of \citet{web11}: a
single catalogue listing for
each of the 295 intervening absorbers (141 Keck plus 154 VLT)
identified along the
131 quasar sightlines probed to-date their MM-based
estimates of $z_\mathrm{abs}$, $\Delta
\alpha/\alpha$, and  $\sigma_\mathrm{obs}$; the last a
 standard
deviation characterizing the uncertainty in each $\Delta
\alpha/\alpha$ estimate arising from explained sources of observational error.  Also
listed
are the J2000 identifier (encoding
right ascension and declination in the equatorial coordinate
system) and redshift, $z_\mathrm{qua}$, of each background
quasar, the telescope on which the corresponding observation was made, and the
$\varepsilon_\mathrm{sys}$ group to which that observation has been assigned for
error analysis---namely, VLT, Keck low contrast (LC), or Keck high contrast (HC).  An additional flag highlights two ``suspect''
absorbers in this catalogue---the
  $z_\mathrm{abs}=1.542$ system toward J00048$-$415728 and the $z_\mathrm{abs}=2.84$ system
  toward J194454+770552.  These objects were omitted from the Webb et al.\
  team's most recent work, and so for consistency
  are omitted from our reanalysis as well.  The quasar dataset thus
  described is available in
  electronic form from Professor Michael
    Murphy's
    \href{http://astronomy.swin.edu.au/~mmurphy/KingJ_12a_VLT+Keck.dat}{homepage}
    at the Swinburne University of Technology.

\subsection{Explained Errors}\label{explained}
An astronomical spectrograph, such as HIRES on
Keck or UVES on the VLT, uses a high-resolution diffraction grating to isolate distinct
wavelength components of the incoming
light, projecting these
 onto a charge-coupled device (CCD) for digital intensity
measurement.  The capture of a raw
astronomical spectrum thus constitutes a photon counting experiment in
which the count in each pixel of the CCD represents a single
draw from a Poisson distribution with mean value set by the
integrated flux intensity of light (within a
narrow wavelength interval) from the observed source \textit{plus} a contaminating contribution from
the background sky and even possibly other non-target emission (e.g.\
stray light from other bright sources in the field of view); all
\textit{modulated} (i.e., non-linearly transformed) by
the inherent response characteristic of the telescope and device.  In
the Webb et al.\ team's analysis multiple such raw spectra
for each quasar were sourced from
the
 Keck and VLT archives and reduced (i.e., combined and
 ``cleaned'') to single sequences of (high quality) absorption line profiles ready for
 modelling.  The reduction process necessarily involves yet further complex
transformations of the raw data (including ``stacking'', ``flat
fielding'', ``bias frame subtraction'', ``wavelength calibration'', ``cosmic ray
 removal'', and
``continuum subtraction''; the essential cleaning and calibration procedures of precision CCD photometry), which ultimately correlate to some
extent both nearby and not-so-nearby pixels in the output
profile sequence.   However, given the almost overwhelming complexity of
accounting directly for each stage of the
 data reduction pipeline in a statistical modelling context the extent of this correlation is assumed negligible---as is standard
practice in the field\footnote{Substantial research efforts are, however,
  underway to develop Bayesian techniques for modelling more
  faithfully the complete
  observational reduction process; see, for example, the recent thesis
  on this topic by \citet{bos11}.}---with each pixel in the reduced
spectrum ultimately treated as independent of all others and assigned an
observational standard deviation, $\sigma_i$.\footnote{For
  further details of this reduction process and the
  rules applied for ``error bar propagation'' therein the interested reader may refer, for instance, to the official
  UVES data reduction cookbook available \href{http://www.eso.org/sci/facilities/paranal/instruments/uves/doc/VLT-MAN-ESO-13200-4033_v83.pdf}{online}.}

As noted earlier the MM method for deriving from the reduced spectrum a $\Delta \alpha / \alpha$
estimate for each intervening absorber involves a complex
multi-parameter fit across its signature absorption lines.  Due to both natural
 (collisional) and Doppler (thermal and/or turbulent) broadening, the
 imprint of each line
 will be typically spread out over multiple resolution
elements and a profile function thus required to characterize its
shape and recover its centroid (i.e., the precise redshift of that component).
The Voigt profile function\footnote{The Voigt profile function
  \citep{ken38,nas06} corresponds in shape to the probability density
  of a continuous random variable defined as the sum of two
  independent random variables of standard forms---a Normal and a
  Cauchy.  For a more detailed description of the use of the Voigt
  profile function in the modelling of astronomical spectra the interested
  reader may refer to (in particular, Appendix A of)
  Michael Murphy's PhD thesis also available
  \href{http://www.phys.unsw.edu.au/astro/research/thesis/MichaelMurphy.pdf}{online}.}---with
three governing  
parameters: redshift, column density, and velocity width; the
latter being constrained jointly (as the kinetic temperature) across all species detected in the
thermal broadening case, but treated as independent in the turbulent case---is well-established as the natural
basis for this procedure.  However, the number of such bases required to accurately
model a particular system, itself a function of the number of distinct ``velocity components'' present in
the intervening cloud, is of course \textit{a priori} unknown.

Thus, the Webb et al.\ team chose to fit each absorber sequentially, starting with a
single Voigt profile function for each line and adding
components until a ``statistically
acceptable'' solution was achieved \citep{kin12}; with the benchmark for acceptability
characterized by a $1/\sigma_i^2$-weighted sum of squared residuals (WSSR) approximately equal to
the number of pixels fit minus the number of free parameters
(dubbed, ``the degrees of freedom'', in their analysis though the
fitted model is in fact
non-linear).  The known limitations of the WSSR as a stopping
criterion for model selection (cf.\
\citealt{mal95,and10}) may well contribute to the
unexplained errors of the quasar dataset discussed below.  Of those Voigt profile models (both thermal and turbulent) tested up
to this maximum number
of components the model with minimum Akaike Information Criterion
(corrected for small $n$) was then selected as the best choice for
\textit{refitting} with $\Delta \alpha / \alpha$ (now) a free parameter.  Having
located (via a downhill gradient-type search)
the WSSR-minimizing (likelihood-maximizing) solution to
this final model the
Webb et al.\ team estimate the magnitude of their explained error
term, $\sigma_\mathrm{obs}$, from the local curvature of the likelihood
surface in $\Delta \alpha / \alpha$; adopting the Normal approximation, $\varepsilon_\mathrm{obs} \sim
\mathcal{N}(0,\sigma_\mathrm{obs}^2)$.  The validity of this
approximation to the shape of the likelihood surface in the stated context has been confirmed via Markov Chain Monte
Carlo (MCMC) exploration, as described by \citet{kin10}.

\subsection{Unexplained Errors}\label{unexplained}
As mentioned in the Introduction this estimate of the uncertainty in $\Delta \alpha
/\alpha$ derived during profile fitting and governed by the (photon counting) realization noise of the
observed spectrum has been declared
 insufficient by the Webb et al.\ team to explain the apparent noise level in the
quasar dataset.  Numerous possible reasons for this have been
proposed and discussed at
length by \citet{mur03} and \citet{kin12}.  These may be
divided broadly into two classes: (\textsc{i}) methodological
limitations of the fitting procedure rendering inadequate the adopted
$\varepsilon_\mathrm{obs}$ distribution, or (\textsc{ii}) hitherto unmodelled sources of
additional noise (possibly systematic).  From the first class we note the neglected impact of
structural
uncertainty in the profile fit with regard to the choice of line
broadening mechanism and number of velocity components to include;
with perhaps a tendency toward underestimation of the latter owing
to the known pitfalls of the WSSR stopping
criterion adopted.  From the second
class we have the chance
blending of lines
from absorbers at different redshifts, and/or run-to-run errors in the
wavelength calibration solution \citep{kin12}.  Dedicated investigations of
the latter
by \citet{gri10} and \citet{whi10} have exposed the presence of intra-order
velocity shifts against iodine cell reference spectra imaged on both the
HIRES and UVES spectrographs of
sufficient magnitude to potentially impact on estimates of the fine structure
constant; see also the novel asteroid-based calibration study of \citet{mol08}.

To account for the inadequacy of $\varepsilon_\mathrm{obs}$ in their
analysis the Webb et al.\ team suppose the existence of a
Normally-distributed error term, $\varepsilon_\mathrm{sys} \sim \mathcal{N}(0,\sigma_\mathrm{sys}^2)$,
adding strictly unbiased (zero mean and mode) noise to their $\Delta\alpha/\alpha$ estimates \citep{web11,kin12,ber12}.
For each of their monopole and/or dipole hypotheses tested against the quasar dataset the authors constrain
the magnitude (viz.\ standard deviation) of this error term, $\sigma_\mathrm{sys}$, through a Least Trimmed Squares
(LTS) procedure \citep{rou84} in which both its estimate and the best-fitting
set of hypothesis parameters
(under a WSSR-based likelihood function including this additional error term)
are refined in an iterative manner.  For
the specific monopole+$r(z)$-dipole hypothesis examined herein (Equation \ref{dipoleeqn} below) the
authors estimate $\sigma_\mathrm{sys} \approx 0.858 \times 10^{-5}$
for their VLT sub-sample (or $\varepsilon_\mathrm{sys}$ group), $\sigma_\mathrm{sys} \approx 1.630 \times
10^{-5}$ for the sub-sample of their Keck absorbers with high contrast
observations (the Keck HC $\varepsilon_\mathrm{sys}$ group), and $\sigma_\mathrm{sys} \approx 0$ for the remainder of their Keck absorbers with low contrast
observations (the Keck LC $\varepsilon_\mathrm{sys}$ group).

Hence, under the Webb et al.\ team's final model for both their explained and
unexplained errors the total uncertainty in each
$\Delta \alpha / \alpha$ estimate is treated as $\varepsilon_\mathrm{tot} \sim
\mathcal{N}(0,\sigma_\mathrm{tot}^2)$ where
$\sigma_\mathrm{tot}^2=\sigma_\mathrm{obs}^2+\sigma_\mathrm{sys}^2$
(with $\sigma_\mathrm{obs}$ unique to each absorber and
$\sigma_\mathrm{sys}$ shared across each $\varepsilon_\mathrm{sys}$ group).  After
outlining for reference the mathematical form of the monopole+$r(z)$-dipole hypothesis
 below we proceed to examine the $\sigma_\mathrm{obs}$- and $\sigma_\mathrm{tot}$-scaled residuals about its best-fit solution
 from the \citet{kin12} study
  as a preliminary (and largely qualitative) evaluation of this simple
  error treatment.

\subsection{Mathematical Form of the Dipole}\label{dpmod}
The monopole+$r(z)$-dipole hypothesis considered by \citet{kin12} and
\citet{ber12} takes the functional form:
\begin{equation}\label{dipoleeqn}
\Delta \alpha /
\alpha{}_{\mathrm{mod}|\bm{x}_i,\bm{\theta}_m} = m + B \times r(z_i)
\cos (\phi)[\mathrm{ra}_i,\mathrm{dec}_i,\mathrm{ra}_d,\mathrm{dec}_d]
\end{equation}
with $\bm{x}_i=\{\mathrm{ra}_i,\mathrm{dec}_i,r(z_i)\}$ the vector of
explanatory variables for the $i$th
absorber, $\bm{\theta}_m=\{
m,B,\mathrm{ra}_d,\mathrm{dec}_d\}$
a vector of input model parameters, and $\cos (\phi)[\cdot]$ a
function returning the cosine of angular
separation between the observational sightline and dipole
vector.  Given right ascension in hours and declination
in degrees the latter may be computed as:
\begin{eqnarray}
\cos (\phi) = & \sin(\mathrm{dec}_i)\sin(\mathrm{dec}_d) +& \\ \nonumber &\cos(\mathrm{dec}_i)\cos(\mathrm{dec}_d)\cos(15\times[\mathrm{ra}_i-\mathrm{ra}_d]).&
\end{eqnarray}
The lookback distance (cf.\ \citealt{hog99}) for each absorber, $r(z_i)$, under the standard cosmological model
(here we adopt $\Omega_M = 0.3$, $\Omega_\Lambda = 0.7$, and $H_0 =
70$ km s$^{-1}$ Mpc$^{-1}$) is given (in units of GLyr; Giga-Light-years) by:
\begin{equation}
r(z_i) = 13.98 \times \int_0^{z_i}
\left[(1+z^\prime)\sqrt{0.3\times (1+z^\prime)^3+0.7}\right]^{-1}
dz^\prime .
\end{equation}
For reference, we note that the monopole-only null hypothesis corresponds simply to the case of
Equation \ref{dipoleeqn} with $B=0$---that is, $\Delta \alpha
/ \alpha{}_{\mathrm{mod}|\bm{\theta}_m=\{m\}} = m$ (a non-zero constant); with $\Delta \alpha
/ \alpha{}_{\mathrm{mod}|\bm{\theta}_m=\{m=0\}} = 0$ its strict null counterpart.

\subsection{Possible Non-Normality of Residuals}\label{nonnorm}
\begin{figure*}
\vspace{0pc}
\includegraphics[width=0.8175\textwidth]{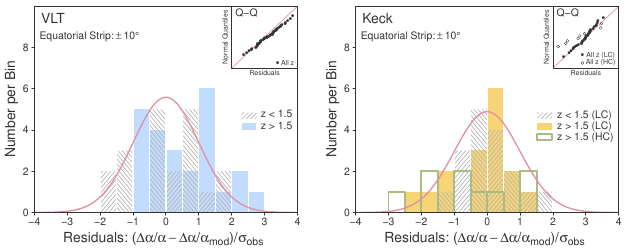}
\includegraphics[width=0.8175\textwidth]{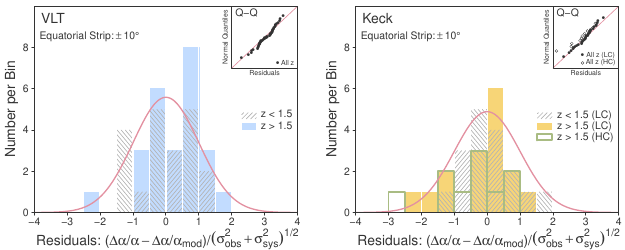}
\vspace{+0.05cm}\caption{Motivation for our consideration of non-Normal error terms.
  Each panel contains a histogram of the residuals, $[\Delta \alpha /
  \alpha - \Delta \alpha /
  \alpha{}_\mathrm{mod}]$, with respect to the best-fit parameterization of the monopole+$r(z)$-dipole
  hypothesis of \citet{kin12}, scaled by the standard
  deviation of the raw
  observational error term only in the top row, $\sigma_\mathrm{obs}$,
  and that of the total error term adopted by Webb et al., $(\sigma_\mathrm{obs}^2+\sigma_\mathrm{sys}^2)^{1/2}$, in the
  bottom row.  To limit any sensitivity to the precise
  functional form adopted for  the dipole we include only those systems lying
  within $\pm10^\circ$ of its equator (VLT sources on the
  left; Keck on the right), and we further subdivide each sample by redshift
  at $z=1.5$.  We draw the reader's attention to
   departures in a number of these histograms from
  the standard Normal form that would be expected if the
  measurement errors were governed entirely by
  Normal terms of the magnitude assumed in each row; in particular, the bimodal split
  about zero in the VLT residuals, and the negative skewness in the Keck
   residuals.  We further emphasise these departures from Normality via
the Q-Q (Normal vs.\ Residual) plots shown as insets in the top-right
corner of each panel; the range here (not marked) is exactly -3 to 3 on each axis. }
\label{kingerrors}
\end{figure*}

In the top row of Figure \ref{kingerrors} we present histograms of the residuals about
the best-fit (i.e., $1/\sigma_\mathrm{tot}^2$ WSSR-minimizing) solution to the monopole+$r(z)$-dipole hypothesis reported
by
\citet{kin12}, scaled by the magnitude
of each absorber's raw observational error term,
$\sigma_\mathrm{obs}$, and subdivided by
telescope (or, more precisely, $\varepsilon_\mathrm{sys}$ group) and
redshift.  To limit any sensitivity to the precise functional form for
the dipole hypothesis 
assumed herein we include only those systems lying within
$\pm$10$^\circ$ of its equator where the contribution of the dipole
component is smaller than or equal to the magnitude of the fitted  monopole
component ($|m|=0.187\times10^{-5}$).  If only the explained uncertainties (cf.\ Section \ref{explained})
represented by $\varepsilon_\mathrm{obs} \sim
\mathcal{N}(0,\sigma_\mathrm{obs}^2)$
were in
operation here (and the monopole+$r(z)$-dipole hypothesis both true and
well fit), then the underlying
distributions of these scaled residuals should, of course, themselves be standard
Normal.  However, this seems unlikely to be the case (as acknowledged
by the Webb et al.\ team's invocation of the unexplained error term) for either the
VLT residuals---which exhibit an unusual bimodality---or the Keck
residuals---which exhibit a strong negative skewness. These departures
from the standard Normal are 
further emphasized by the Q-Q plots shown in the top-right inset of
each panel.

As described in Section \ref{unexplained}, under the Webb et al.\ team's model for the unexplained uncertainties
in the quasar dataset
the \textit{total} error on each observation remains zero mean Normal,
$\varepsilon_\mathrm{tot} \sim \mathcal{N}(0,\sigma_\mathrm{tot}^2)$, but takes
a broader standard deviation (at least in the case of the VLT and Keck HC
$\varepsilon_\mathrm{sys}$ groups) of
$\sigma_\mathrm{tot}=(\sigma_\mathrm{obs}^2+\sigma_\mathrm{sys}^2)^{1/2}$.  Rescaling the
VLT and Keck HC residuals accordingly we recover those histograms shown
in the bottom row of Figure \ref{kingerrors}.  While the scaled
residuals now bear a greater similarity to the standard
Normal, from
a skeptical perspective the
rough agreement seen here cannot be considered reassuring,
especially given the lack
of any strong justification for the supposed form of
$\varepsilon_\mathrm{tot}$. By including below a number of plausible
parametric alternatives to the
Webb et al.\ team's nominal error model in
our reanalysis of the quasar dataset we are able to gauge the
influence of the assumed form for the unexplained errors on the
apparent dipole significance (cf.\ Sections
\ref{generativemodels} to \ref{marginallikelihoods}).

\subsection{Twice-Observed Absorbers}\label{twiceobs}
As noted earlier although the Keck and VLT telescopes cover generally well-separated regions of sky---the Hawaii-based Keck being largely
restricted to targets in the Northern hemisphere, and the Chile-based
VLT restricted to targets in the Southern hemisphere---there
nevertheless appears in the Webb et al.\ dataset an eight quasar intersection of sightlines common
to both (evident in Figure \ref{kingdipole} as the star-shaped
symbols arising from the overlap of the
up-triangle and down-triangle symbols used respectively for the VLT and Keck
datapoints).  Amongst the corresponding population of intervening absorbers thus
probed there are eleven twice-observed; that is, detected at matched
spectroscopic redshifts in the
spectra from both instruments (which feature quite different
observational selection functions---owing to the unique resolution and
wavelength coverages of each---and hence do not necessarily return
\textit{identical} sets of absorbers).  The pairwise comparison of Keck and VLT $\Delta \alpha / \alpha$
estimates for these twice-observed systems offers a ``first-order'' test of
the unbiased error hypothesis questioned by Chad Orzel and others skeptical
of the Webb et al.\ team's conclusions.

The eleven scaled differences, $[\Delta \alpha / \alpha_\mathrm{Keck}-\Delta \alpha /
\alpha_\mathrm{VLT}]/(\sigma_\mathrm{tot,Keck}^2+\sigma_\mathrm{tot,VLT}^2)^{1/2}$, so recovered form the ordered
list: $\{ -1.64, -1.40, -0.97, -0.75, -0.56, -0.52,\\ -0.34, -0.09,
0.17, 0.27, 0.30 \}$ (rounded to the 2nd decimal place).  The predominance of negative
differences observed here (eight of the eleven) hints at the
presence of systematic biases in the $\Delta \alpha / \alpha$ values
recovered from one or both of these telescopes; and indeed a Wilcoxon
signed-rank test \citep{wil45} allows rejection of the zero mean
difference 
hypothesis at the 5\% significance 
level ($p=0.017$).  Though admittedly not of overwhelming power due to the small
sample size available this result nevertheless serves as
further motivation (in addition to the alignment of the alleged
dipole's equator along the coverage overlap of the Keck and VLT telescopes) for the biased error models we consider in our Bayesian
reanalysis of the quasar dataset
 below.

\section{Generative Models}\label{generativemodels}
Having established our motivation for a Bayesian reanalysis of the
quasar dataset, and for exploring in this context both unbiased and
biased error terms, in the Introduction and Section
\ref{quasardataset}, respectively, we now proceed to outline formally the
 generative models and priors required for this endeavour.

\subsection{Error Models}\label{errormodels}
In the interests of thoroughness we explore four alternative, parametric
forms for the total error term operating in each
$\varepsilon_\mathrm{sys}$ group; each candidate form allowing a quite
different distribution for the unexplained error component.  For reference we denote
these, \textsc{\bf{A}},
\textsc{\bf{B}}, \textsc{\bf{C}}, and \textsc{\bf{D}}, respectively, and define as
follows:
\[(\textsc{\bf{A}}) \
\varepsilon_\mathrm{tot}
\sim
\mathcal{N}(\beta_\mathrm{sys},\sigma_{\mathrm{obs}}^2+\sigma_\mathrm{sys}^2),
\]
\[(\textsc{\bf{B}}) \
\varepsilon_\mathrm{tot}
\sim
\mathcal{V}(\beta_\mathrm{sys},\sigma_{\mathrm{obs}},\sigma_\mathrm{sys}),
\ [\mathcal{V}
\mathrm{\ the\ Voigt\ distribution}]\]
\[ (\textsc{\bf{C}})\
\varepsilon_\mathrm{tot} \sim
\mathcal{N}_\mathrm{skew}(\xi_\mathrm{sys},\frac{v_\mathrm{sys}w_\mathrm{sys}}{\sqrt{\sigma_\mathrm{obs}^2+w_\mathrm{sys}^2}},\sigma_\mathrm{obs}^2+w_\mathrm{sys}^2)
, \]
\[(\textsc{\bf{D}})\
  \varepsilon_\mathrm{tot}
   \sim \left\{ \begin{array}{ll}
 &\mathcal{N}(-\gamma_\mathrm{sys}\sigma_\mathrm{obs},\sigma_\mathrm{obs}^2)\mathrm{\
 with\ probability\ }\eta_\mathrm{sys},\\
 & \mathcal{N}(\gamma_\mathrm{sys}\sigma_\mathrm{obs},\sigma_\mathrm{obs}^2)\mathrm{\
   otherwise}.\end{array}\right.
\]

Error form \textsc{\bf{A}} corresponds, of course, to the scenario of a
Normally-distributed source of systematic measurement noise,
$\varepsilon_\mathrm{sys} \sim \mathcal{N}(\beta_\mathrm{sys},\sigma_\mathrm{sys}^2)$, operating in addition
to, and independently of, the
explained error, $\varepsilon_\mathrm{obs} \sim \mathcal{N}(0,\sigma_\mathrm{obs}^2)$.  At $\beta_\mathrm{sys}=0$
this equates to the strictly unbiased (i.e., zero mean
and mode) error term adopted by
\citet{kin12}; and we therefore consider this limiting case the
\textit{nominal} error model ($\textsc{\bf{A}}_0$) for our analysis, treating the more
general $\beta_\mathrm{sys}
\neq 0$ case as the default \textit{skeptical} model.

Error form
\textsc{\bf{B}}, on the other hand, corresponds to the (undesirable) scenario
of a \textit{heavy-tailed} source of Cauchy-distributed
measurement noise with location, $\beta_\mathrm{sys}$, and scale,
$\sigma_\mathrm{sys}$, i.e., $\varepsilon_\mathrm{sys} \sim
\mathrm{C}(\beta_\mathrm{sys},\sigma_\mathrm{sys})$ where
$f_{\mathrm{C}}(x)=1/\pi \times
\sigma_\mathrm{sys}/(\sigma_\mathrm{sys}^2+(x-\beta_\mathrm{sys})^2)$,
operating in addition to, and independently of, the explained error.  As noted in our earlier discussion of absorption
line modelling (Section \ref{explained}), the Voigt profile function
(cf.\ \citealt{ken38,nas06}), $\mathcal{V}$, provides an
analytical expression for the density of
such a distribution arising from the convolution of a
Cauchy and a Normal.  Namely,
\[f_{\mathcal{V}}(x)=\frac{\mathrm{Re}[\omega(z)]}{\sigma_\mathrm{obs}\sqrt{2\pi}},
\ z=\frac{(x-\beta_\mathrm{sys})+i
  \sigma_\mathrm{sys}}{\sigma_\mathrm{obs}\sqrt{2}}, -\infty < x < \infty,
\] where $\omega(\cdot)$ in the above denotes the complex
error (or ``Faddeeva'') function.  By allowing, through error form $\textsc{\bf{B}}$,
for the
possibility of
heavy-tailed behaviour in the unexplained error
term we help to ensure the robustness of our conclusions under the
Normal error form 
(\textsc{\bf{A}}) against the possible presence of extreme outliers in the
quasar
dataset.  Though the mean remains undefined for the Cauchy, and
thus the Voigt as well, we note that $\beta_\mathrm{sys} = 0$
nevertheless gives
a zero mode case 
($\textsc{\bf{B}}_0$), which we refer to as ``unbiased'' for our purposes
and which we examine separately to the more general
(skeptical) $\beta_\mathrm{sys} \neq 0$ case.

Error form \textsc{\bf{C}} supposes instead a
 source of \textit{skew Normal} noise, $\varepsilon_\mathrm{sys} \sim
 \mathcal{N}_\mathrm{skew}(\varepsilon_\mathrm{sys},v_\mathrm{sys},w_\mathrm{sys}^2)$, operating in addition to, and
 independently of, the explained
error.  A tendency toward bias in the
wavelength calibration solution for a given $\varepsilon_\mathrm{sys}$ group
may be one explanation for such an asymmetric distribution in the
uncertainties of the quasar dataset.
We note here for reference the definition of the skew Normal with normalized
shape parameter, $v$,
scale, $w$, and location parameter, $\xi$, in terms of the
standard Normal density, $\phi(\cdot)$, and distribution function, $\Phi(\cdot)$:
\[
f_{\mathcal{N}_\mathrm{skew}}(x) = \frac{2}{w}
\phi\left(\frac{x-\xi}{w}\right)
\Phi\left(\frac{v (x-\xi)}{w\sqrt{1-v^2}}\right),\ -\infty < x < \infty,
\]
with $-1 < v < 1$ and $w > 0$.  We note also that the sum of a (zero
mean) Normal,
$\mathcal{N}(0,\sigma_\mathrm{sys}^2)$, and a skew Normal,
$\mathcal{N}_\mathrm{skew}(\xi_\mathrm{sys},v_\mathrm{sys},w_\mathrm{sys}^2)$,
is again skew Normal, with parameters as specified in our definition of
 error form \textsc{\bf{C}} above.
 This may be verified, of course, by reference to the characteristic
function of the skew Normal distribution; according to \citet{pew00}, $\Psi_{\mathcal{N}_\mathrm{skew}(\xi,v,w^2)}(t) =
\exp(i\xi t-w^2t^2/2)\{ 1 + i\tau(vwt) \}$ where
$\tau(x)=\int_0^x\sqrt{2/\pi}\exp(u^2/2)du$ for $x >0$ and
$\tau(-x)=-\tau(x)$.  We consider once again a so-called ``unbiased''
(here zero
mode, but \textit{not} mean) case ($\textsc{\bf{C}}_0$), constructed by solving numerically
for
$\xi_\mathrm{mode}$ for each $\sigma_\mathrm{obs}$ and
$\{v_\mathrm{sys},w_\mathrm{sys}\}$ parameter pairing, in
addition to the skeptical case of free $\xi_\mathrm{sys}$.

Finally, error form
\textsc{\bf{D}} corresponds to a scenario of
systematic mis-estimation in which $\Delta \alpha
/ \alpha$ is alternately \textit{under}-shot or \textit{over}-shot by
$\gamma_\mathrm{sys}\times\sigma_\mathrm{obs}$, with probability
$\eta_\mathrm{sys}$ of the former (and $1-\eta_\mathrm{sys}$ of the
latter).  Such a bimodal error distribution might well arise
as the result of routine mis-identification of the dominant line broadening
mechanism (turbulent or thermal) or the number of velocity
components present when modelling the observed absorption profile (cf.\ Section \ref{explained}); and was suggested
by the distribution of residuals around the equator of the proposed
dipole for the VLT sample (examined in
Section \ref{nonnorm}).  In the interests of thoroughness,
though perhaps not terminological consistency,
we also consider an ``unbiased'' (here zero mean, but not mode) case
of this error form ($\textsc{\bf{D}}_0$) with fixed $\eta=0.5$.

\begin{figure*}
\includegraphics[width=0.815\textwidth]{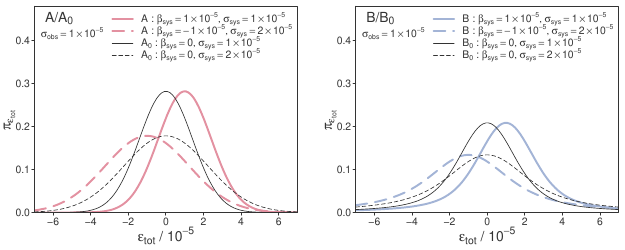}
\includegraphics[width=0.815\textwidth]{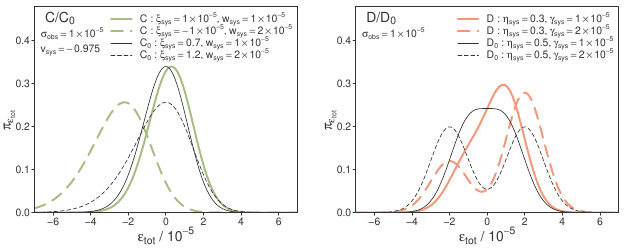}
\vspace{-0.0cm}\caption{Illustration of the four candidate forms 
  ($\textsc{\bf{A}}$, $\textsc{\bf{B}}$, $\textsc{\bf{C}}$, and
  $\textsc{\bf{D}}$) proposed here for modelling 
  the total error term, $\varepsilon_\mathrm{tot}$, of the quasar
  dataset.  We specify an arbitrary standard deviation of 
  $\sigma_\mathrm{obs}=1\times10^{-5}$ for the explained uncertainty component,
$\varepsilon_\mathrm{obs}\sim\mathcal{N}(0,\sigma_\mathrm{obs}^2)$,
 in these plots,   
and we indicate the diversity of densities thereby admitted under each 
corresponding total
error term by
tracing the output for a variety of
selected 
inputs to its governing parameters (detailed in the top-right legend of each
panel).  For clarity, the ``unbiased'' cases of the forms shown here ($\textsc{\bf{A}}_0$, $\textsc{\bf{B}}_0$, $\textsc{\bf{C}}_0$, and
  $\textsc{\bf{D}}_0$)  are clearly distinguished (via a reduced line
  thickness) from their general (skeptical) counterparts.  }
\label{emodels}
\end{figure*}

In Figure \ref{emodels} we illustrate the nature and diversity of all
four candidate forms ($\textsc{\bf{A}}$, $\textsc{\bf{B}}$,
$\textsc{\bf{C}}$, and $\textsc{\bf{D}}$)
for the total error term, $\varepsilon_\mathrm{tot}$, under a
variety of arbitrary choices for their governing parameters.  As in 
earlier studies by the Webb et al.\ team, each of the three
$\varepsilon_\mathrm{sys}$ groups of the quasar dataset---namely, the VLT sample, the low
contrast (LC) Keck sample, and the high contrast (HC) Keck
sample---is treated separately here for error analysis and
assigned one of the above.  Throughout we employ the notation,
$\textsc{\bf{AAA}}$,
$\textsc{\bf{B}}_0$$\textsc{\bf{B}}_0$$\textsc{\bf{B}}_0$,
$\textsc{\bf{C}}$$\textsc{\bf{A}}$$\textsc{\bf{D}}$ and so on, to
indicate which error form has been assigned to each---Keck LC, Keck
HC, and VLT, in that order.  We refer specifically to the
combination, 
$\textsc{\bf{A}}\textsc{\bf{A}}\textsc{\bf{A}}$, as our default
skeptical model and to the combination, $\textsc{\bf{A}}_0\textsc{\bf{A}}_0\textsc{\bf{A}}_0$, as
the nominal error model.  With each candidate form admitting two or three governing parameters we must therefore
specify a prior density on up a total of nine parameters
for a complete error model.  

\subsection{Error Model Priors}\label{hyperpriors}
It is widely acknowledged (at least in the statistical literature)
that the posterior Bayes factor for Bayesian model selection (BSM) can
be remarkably sensitive to the chosen priors on the internal parameters
of the various models under consideration (cf.\ \citealt{kas95}).  In
particular, unlike in the case of general Bayesian inference the
interpretation of improper priors is inevitably ill-defined in the model
selection context, and the influence of the prior choice cannot simply be
assumed to grow negligble with the data volume, $n$, any faster than
an $\mathcal{O}(n^{-1})$ rate.  Hence, both the justification of one's
prior selections and a subsequent analysis of the problem-specific
prior-sensitivity are essential ingredients of rigorous BSM; the
former we present directly below and the latter in Section \ref{marginallikelihoods}.

In constructing
priors on the governing parameters of our error models we are guided
by the general observation
that the standard deviation of the unexplained error term must be of
roughly the same order as the explained error term in this dataset
($\overline{\sigma_\mathrm{obs}} \approx 2 \times 10^{-5}$); if it were
very much smaller it would presumably have gone unnoticed, while if it
were very much larger the Webb et al.\ team would have been unlikely
to have proceeded to hypothesis testing without a better understanding of its
origin.  Equivalently, its bias (under the skeptical interpretation)
should be of comparable magnitude (in $\Delta\alpha/\alpha$) to that of
the alleged dipole.  Choosing independent Normal densities to
describe our prior beliefs regarding the  $\beta_\mathrm{sys}$ and
$\sigma_\mathrm{sys}$ of our default skeptical error form
($\textsc{\bf{A}}$) we reduce the problem of prior specification to
one of hyperparameter choice \citep{chi01}; and in light of the above
we nominate simply,
$\pi(\beta_\mathrm{sys})\sim\mathcal{N}(0,\sigma_q^2)$ with (hyperparameter) $\sigma_q = 0.5\times10^{-5}$ and
$\pi(\sigma_\mathrm{sys})\sim\mathcal{N}_\mathrm{half}(0,\sigma_p^2)$
with (hyperparameter) $\sigma_p = 2\times10^{-5}$. (Here
$\mathcal{N}_\mathrm{half}$ denotes the half-Normal distribution,
chosen so as to keep the standard deviation parameter, $\sigma_\mathrm{sys}$, strictly positive.)

To promote
fairness in the marginal likelihood-based comparison of each
model--hypothesis pairing we have attempted to make our priors on the
parameters of the
remaining three error forms as similar as possible to those of form
$\textsc{\bf{A}}$.  Thus, for error
form $\textsc{\bf{B}}$ we suppose 
$\pi(\beta_\mathrm{sys})\sim\mathcal{N}(0,[0.5\times10^{-5}]^2)$ for the mode of the underlying Cauchy, but allow a
(slightly) larger range of
$\pi(\sigma_\mathrm{sys})\sim\mathcal{N}_\mathrm{half}(0,[\sqrt{2\log(2)}\times2\times10^{-5}]^2)$
on its scale parameter to match our effective prior on the full width at half maximum
(FWHM) of this error form to that of $\textsc{\bf{A}}$
(the standard
deviation being undefined for the Cauchy, and thus not available
for comparison).  For error form $\textsc{\bf{C}}$ we begin by setting our prior on the
\textit{unnormalized} skew parameter,
$v_\mathrm{sys}/\sqrt{1-v_\mathrm{sys}^2}$, of this distribution to a
generous, $\mathcal{N}(0,[10]^2)$; which transforms to the following
density on the \textit{normalized} skew parameter, $v_\mathrm{sys}$,
with $a=10$:
\[
f_{\pi(v_\mathrm{sys})}(x) = \frac{1}{\sqrt{2 \pi a^2}}
\exp\left[\frac{-x^2}{2 a^2 (1-x^2)}\right]\frac{1}{(1-x^2)^{3/2}},
\]
for $-1 < x < 1$ (zero otherwise).
We then match our prior on the standard deviation of the unexplained
error term in $\textsc{\bf{C}}$ to that of $\textsc{\bf{A}}$ by
setting $\pi(w_\mathrm{sys}|v_\mathrm{sys}) \sim \mathcal{N}_\mathrm{half}(0,[(1-\frac{2}{\pi}v_\mathrm{sys}^2)^{-1/2}\times 2
\times 10^{-5}]^2)$; and likewise to match our priors on the
corresponding means we choose,
$\pi(\xi_\mathrm{sys}|v_\mathrm{sys},w_\mathrm{sys}) \sim
\mathcal{N}(-v_\mathrm{sys}w_\mathrm{sys}\sqrt{\frac{2}{\pi}},[0.5\times10^{-5}]^2)$.
Finally, for error form $\textsc{\bf{D}}$, for which no precise
matching of this nature is possible, we simply adopt 
$\pi(\gamma_\mathrm{sys}) \sim \mathcal{N}(0,[0.5\times10^{-5}]^2)$ and 
$\pi(\eta_\mathrm{sys})\sim\mathrm{Beta}_\mathrm{half}(15,15)$ (i.e.,
``folded'' about 0.5 [$f^\ast_{0 < x < 0.5} \mathrm{\ or \ } f^\ast_{0.5 < x < 1} = 2 f_{0 < x < 1}$], favouring a near-symmetric
$\varepsilon_\mathrm{sys}$ over a  markedly asymmetric one.  For the typical absorber with explained error,
$\sigma_\mathrm{obs}\approx 2\times10^{-5}$, this gives a prior
expectation on the bias of the unexplained error term in each
$\varepsilon_\mathrm{sys}$ group comparable to
that of our form $\textsc{\bf{A}}$
benchmark.

Important to note is that we have avoided Uniform priors at the above
specification stage; the reasons behind this are two-fold:
\textsc{(i)} although computationally convenient, the implication of a
Uniform prior choice---that one believes all parameter values
within a precise parameter range equally plausible but anything
outside this range entirely implausible---is clearly not justifiable
for any of these parameters; and \textsc{(ii)} provided the range
hyperparameters of a Uniform prior are made sufficiently wide as to
encompass the bulk of posterior mass, then their further manipulation
has negligble impact on the posterior fit but a profound impact on the
corresponding Bayes factor, meaning that such priors can all too easily be manipulated
by the unscrupulous practitioner to artificially favour or disfavour
some 
particular model.

\subsection{Hypothesis Priors}\label{priors} We apply similar principles with regard to constructing priors on the input parameters, $\bm{\theta}=\{
m,B,\mathrm{ra}_d,\mathrm{dec}_d\}$, of the Webb et al.\ team's
proposed monopole+$r(z_i)$-dipole
hypothesis for the
apparent cosmic variation of
$\Delta\alpha/\alpha$ (see Equation \ref{dipoleeqn} in Section \ref{dpmod}).  For the strength of the monopole
term, $m$, we adopt a zero mean Normal prior with standard deviation,
$0.5 \times 10^{-5}$, and for the strength of the dipole term, $B$, an exponential
prior with rate, $1/[0.5\times10^{-5}]$.  That is,
$\pi(m)\sim\mathcal{N}(0,[0.5\times10^{-5}]^2)$ and
$\pi(B)\sim\mathrm{Exp}(1/[0.5\times10^{-5}])$.  Supposing \textit{a
  priori} that the
dipole vector might point anywhere on the celestial sphere with equal
probability we assign the appropriate uniform priors to both $\mathrm{ra}_d$ and the
sine of $\mathrm{dec}_d$.  Thus, $\pi(\mathrm{ra}_d) \sim
\mathrm{U}(0,24)$ and $\pi(\sin(\mathrm{dec}_d)) \sim \mathrm{U}(-1,1)$, i.e., $\pi(\mathrm{dec}_d) \sim
\cos(\mathrm{dec}_d)/2$. 

For consistency with the above we also suppose
$\pi(m)\sim\mathcal{N}(0,[0.5\times10^{-5}]^2)$ for the strength of
the monopole in the monopole-only version of the null
hypothesis; recall though, our strict null
hypothesis is that $\Delta \alpha / \alpha$
remains everywhere zero.

\subsection{Likelihood Function}
With any stochastic variation in the $\Delta \alpha / \alpha$
estimates of the quasar dataset assumed to arise solely
from the combination, $\varepsilon_\mathrm{tot}$, of the explained and unexplained error terms
described by the error models of Section \ref{errormodels}---and with the
realization of this measurement error for any given absorber assumed
independent of all others---the likelihood function for the observed
data,
$\bm{y}=\{ \Delta\alpha/\alpha{}_i: i=1,\ldots,293\}$
with covariates represented as $\bm{x}_i=\{\mathrm{ra}_i,\mathrm{dec}_i,r(z_i)\}$, corresponding to
a given $\Delta \alpha/\alpha{}_\mathrm{mod}$ and
$\varepsilon_\mathrm{tot}$ pairing, $M$, may be written  as:
\begin{eqnarray} \nonumber
&L(\bm{y}|\bm{\theta}=\{\bm{\theta}_m,\bm{\theta}_e\},M,\bm{x}_i) =& \\
&\prod_{i=1}^{293}
f_\mathrm{\varepsilon_\mathrm{tot}|\bm{\theta}_e}(\Delta\alpha/\alpha{}_i -
\Delta\alpha/\alpha{}_{\mathrm{mod}(M)|\bm{x}_i,\bm{\theta}_m}).&
\end{eqnarray}
Here $\bm{\theta}_m=\{
m,B,\mathrm{ra}_d,\mathrm{dec}_d\}$ (dipole) [or $\{
m\}$ (monopole null), or $\{m=0\}$ (strict null)] denotes the set of input
hypothesis parameters,
$f_{\varepsilon_\mathrm{tot}|\bm{\theta}_e}$
the $\varepsilon_\mathrm{tot}$ probability density for the
particular 
$\varepsilon_\mathrm{sys}$ group to which the $i$th absorber belongs, and
$\bm{\theta}_e$ the vector of between six and nine parameters required
for a complete error model.

\subsection{Posteriors}\label{posteriors}
For each combination of error model and hypothesis tested here (see Section
\ref{marginallikelihoods} below for a complete enumeration) we have constructed a Markov Chain
Monte Carlo (MCMC) sample of 10,000 draws from the posterior ($\beta=1$), plus a
further 10,000 draws from the prior ($\beta=0$) and each of eight tempered,
bridging densities ($\pi(\theta)L(\theta)^\beta$ with $\beta$
logarithmically spaced between $10^{-5}$ and $0.5$), via the method of
``parallel tempering'' (also known as MC$^3$; \citealt{swe86,gey92}).
By allowing probabilistic swaps between chains at different
temperatures the parallel tempering method can greatly improve mixing
relative to the rate of each run separately,
particularly for multimodal posteriors.  For our within-chain moves we
use a symmetric, random walk type proposal; the diagonal variance matrix
of which was initialized to our prior variance
on each parameter and progressively rescaled toward a target acceptance rate of
0.4 during a 5,000 draw burn-in phase.  Although the marginal
likelihoods of each model--hypothesis pairing (which we ultimately derive from
these chains using the RLR technique described in Section
\ref{rlr} below) are in fact the sole quantities of
interest for our Bayesian assessment of the
claimed spatial variation in the quasar dataset we nevertheless review
briefly here the nature of our posterior inferences for key
error model and hypothesis parameters as an aid to understanding
the ensuing results.

\begin{figure*}
\includegraphics[width=0.815\textwidth]{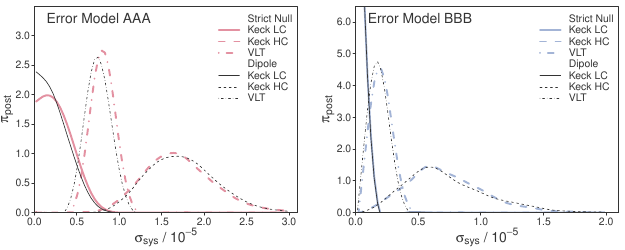}
\includegraphics[width=0.815\textwidth]{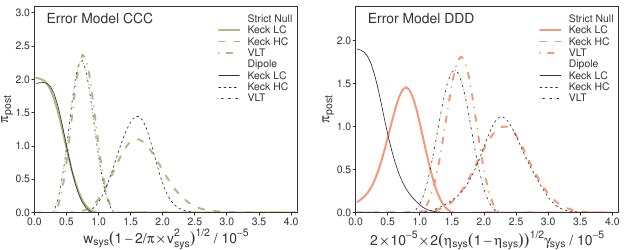}
\vspace{+0.075cm}\caption{Posterior probability densities for the key
  parameters (or combinations thereof) governing the
  underlying error distribution width (viz.\ standard deviation, or half
  width half maximum as appropriate in the $\textsc{\bf{BBB}}$ case) of the $\varepsilon_\mathrm{sys}$ component in
  each of our general (skeptical) error models: $\textsc{\bf{AAA}}$ to
  $\textsc{\bf{DDD}}$ (homogeneous).  In the interests of illustrative
  clarity only those posteriors corresponding to the strict null
 (i.e., $\Delta \alpha / \alpha$ everywhere zero) and dipole
hypotheses are compared here (our results for the monopole-only null
hypothesis being roughly intermediate to these two).  Interestingly, although the inferred
width of the unexplained error term for each $\varepsilon_\mathrm{sys}$
group varies somewhat depending upon the error model adopted, the
relative ordering of the aforesaid---Keck LC, VLT, Keck HC (from smallest to largest)---does not.}
\label{mcmcerrors}
\end{figure*}

\begin{figure*}
\includegraphics[width=0.815\textwidth]{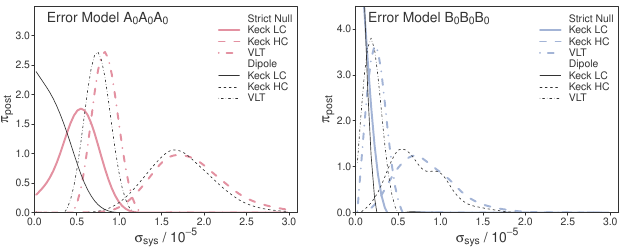}
\includegraphics[width=0.815\textwidth]{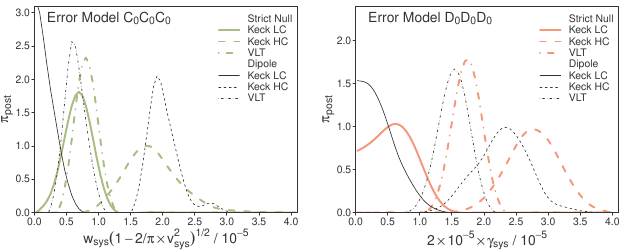}
\vspace{0.075cm}\caption{Posterior probability densities for the key
  parameters (or combinations thereof) governing the
  underlying error distribution width (viz.\ standard deviation, or half
  width half maximum as appropriate in the $\textsc{\bf{B}}_0\textsc{\bf{B}}_0\textsc{\bf{B}}_0$ case) of the $\varepsilon_\mathrm{sys}$ component in
  each of our ``unbiased'' error models:
  $\textsc{\bf{A}}_0$$\textsc{\bf{A}}_0$$\textsc{\bf{A}}_0$ to
  $\textsc{\bf{D}}_0$$\textsc{\bf{D}}_0$$\textsc{\bf{D}}_0$ (homogeneous).  Again, although the inferred
width of the unexplained error term for each $\varepsilon_\mathrm{sys}$
group varies somewhat depending upon the error model---and hypothesis---adopted, the
relative ordering of the aforesaid---Keck LC, VLT, Keck HC (from
smallest to largest)---does not.  As expected the error widths
recovered under the null hypothesis here are generally somewhat
greater than those recovered under the (more flexible) dipole hypothesis.}
\label{mcmcerrors2}
\end{figure*}

In Figures \ref{mcmcerrors} and \ref{mcmcerrors2} we illustrate the marginal posterior
probability densities of those parameters (or combinations thereof) governing the
width (viz.\ standard deviation, or half width half maximum as
appropriate in the case of model $\textsc{\bf{B}}$) resulting from
each error form ($\textsc{\bf{A}}$/$\textsc{\bf{A}}_0$ to $\textsc{\bf{D}}$/$\textsc{\bf{D}}_0$, as described
in Section \ref{errormodels}) applied homogeneously across all
$\varepsilon_\mathrm{sys}$ groups under both the
strict null and dipole hypotheses.  (Our posteriors under the monopole-only null
hypothesis, and for the two non-homogeneous error model combinations we test, $\textsc{\bf{CAA}}$ and
$\textsc{\bf{CAD}}$ [cf.\ Section \ref{marginallikelihoods} below], being roughly intermediate to these representative examples.)  Although the inferred
width of the unexplained error term for each $\varepsilon_\mathrm{sys}$
group varies somewhat depending upon the error model adopted, the
relative ordering of the aforesaid---Keck LC, VLT, Keck HC (from smallest to largest)---does not.  For the ``unbiased'' case of each model in particular (Figure
\ref{mcmcerrors2}: $\textsc{\bf{A}}_0$$\textsc{\bf{A}}_0$$\textsc{\bf{A}}_0$ to $\textsc{\bf{D}}_0$$\textsc{\bf{D}}_0$$\textsc{\bf{D}}_0$) we note that the error widths
recovered under the null hypothesis are generally somewhat greater than
those recovered under the (more flexible) dipole hypothesis---that is,
the
proposed error forms must necessarily ``stretch'' to encompass the
broader spread of residuals in the former.
Interestingly, the Bayesian analysis we present here for the nominal error
model ($\textsc{\bf{A}}_0$$\textsc{\bf{A}}_0$$\textsc{\bf{A}}_0$) favours a non-zero
$\sigma_\mathrm{sys}$ of roughly 0.6$\times10^{-5}$ (posterior mean)
for the Keck LC $\varepsilon_\mathrm{sys}$
group under the strict null hypothesis, in constrast to the
$\sigma_\mathrm{sys} \approx 0$ Least Trimmed Squares (LTS) estimate
reported by \citet{kin12} in their Table 2.  This discrepancy
presumably arises from the deliberate focus on
interquartile width at the expense of (or, purportedly in robustness against) any marked outliers in the latter approach.  Our estimates for the other
$\varepsilon_\mathrm{sys}$ groups under both the null and dipole
hypotheses are nevertheless in fair agreement with their LTS counterparts.

\begin{figure*}
\vspace{0pc}
\includegraphics[width=0.815\textwidth]{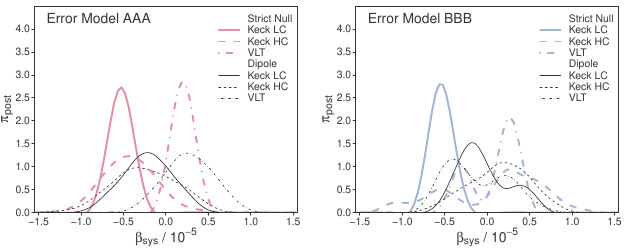}
\includegraphics[width=0.815\textwidth]{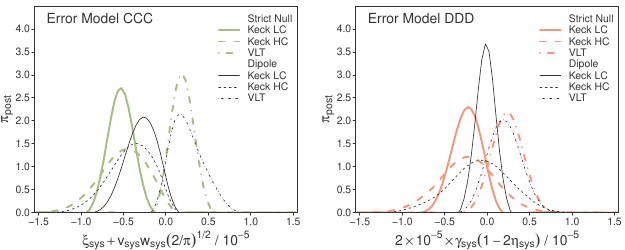}
\vspace{-0.00cm}\caption{Posterior probability densities for the key
  parameters (or combinations thereof) governing the error distribution bias
  (viz.\ offset of the mode, or in case $\textsc{\bf{D}}$ the mean, from zero) in
  each of the biased error models:
  $\textsc{\bf{A}}$$\textsc{\bf{A}}$$\textsc{\bf{A}}$ to
  $\textsc{\bf{D}}$$\textsc{\bf{D}}$$\textsc{\bf{D}}$ (homogeneous).  Once again in the interests of illustrative
  clarity only those posteriors corresponding to our strict null
hypothesis and dipole
model are compared here.  The degree of bias favoured by
the results for each $\varepsilon_\mathrm{sys}$ group varies little
between error models but greatly between hypotheses---as expected, much
larger biases are inferred under the strict null hypothesis than under
the dipole hypothesis, owing to the inherent (though of course only partial) degeneracy between the North--South
dipole and opposing bias scenarios.}
\label{mcmcbetas}
\end{figure*}

In Figure \ref{mcmcbetas} we present the marginal posterior
probability densities of the parameters (or combinations thereof) governing the ``bias''
(viz.\ offset of the mode, or in case $\textsc{\bf{D}}$ the mean, from zero) in each of our
general (skeptical)
error models (applied homogeneously across all $\varepsilon_\mathrm{sys}$ groups) under both the
strict null and dipole hypotheses.  The degree of bias favoured by
these posteriors for each $\varepsilon_\mathrm{sys}$ group varies little
between error models but greatly between hypotheses---as expected, 
larger biases are generally preferred under the strict null hypothesis than under
the dipole hypothesis, owing to the inherent (though of course only partial) degeneracy between the North--South
dipole and opposing bias scenarios.

\begin{figure*}
\vspace{0pc}
\includegraphics[width=0.815\textwidth]{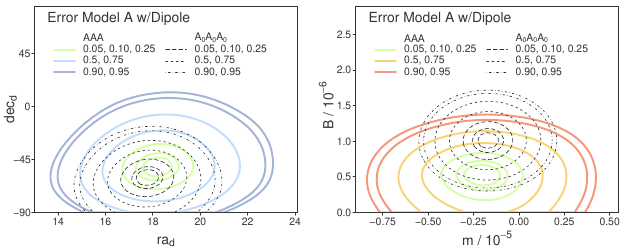}
\includegraphics[width=0.815\textwidth]{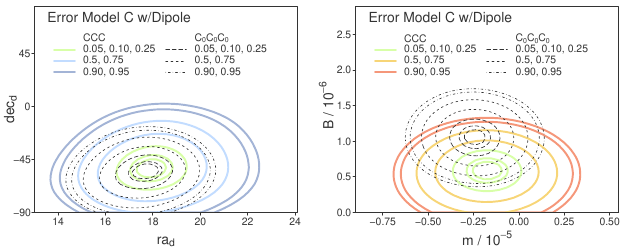}
\vspace{-0.0cm}\caption{Posterior probability densities for the four key
  parameters, $\bm{\theta}_m=\{
m,B,\mathrm{ra}_d,\mathrm{dec}_d\}$, governing the Webb et al.\ team's
proposed monopole+$r(z)$-dipole hypothesis for the apparent spatial
variation of $\Delta \alpha
/ \alpha$ in the fine structure dataset under the following
alternative, example
forms of the unexplained error term:
$\textsc{\bf{A}}$$\textsc{\bf{A}}$$\textsc{\bf{A}}$ and
$\textsc{\bf{A}}_0$$\textsc{\bf{A}}_0$$\textsc{\bf{A}}_0$ (\textit{top
  row}); and
$\textsc{\bf{C}}$$\textsc{\bf{C}}$$\textsc{\bf{C}}$ and $\textsc{\bf{C}}_0$$\textsc{\bf{C}}_0$$\textsc{\bf{C}}_0$ (\textit{bottom row}).  We illustrate
each posterior via a series of smoothed contours (surrounding the
posterior mode and) enclosing increasing
fractions of the (marginal) posterior mass from $0.05$
to $0.95$ in both the $m$--$B$ plane (giving the strengths of the monopole and
dipole components, respectively)
and the $\mathrm{ra}_d$--dec$_{d}$ plane (giving the dipole direction on
the celestial sphere).  While the posteriors under each
error model in the ``unbiased'' case
are both highly concentrated around the maximum likelihood
solutions identified by the Webb et al.\ team under their fitting of
the nominal
error model ($\textsc{\bf{A}}_0$$\textsc{\bf{A}}_0$$\textsc{\bf{A}}_0$)---namely,
$\{\mathrm{ra}_d\approx17.5$ hours, $\mathrm{dec}_d\approx-60$
degrees$\}$ and $\{m\approx-0.2\times10^{-5}$,
$B\approx10^{-6}\}$---in the biased (skeptical) case the corresponding
posteriors are evidently more diffuse
and favour markedly a smaller dipole strength.}
\label{mcmcMB}
\end{figure*}

Finally, for reference we plot in Figure
\ref{mcmcMB} the joint posterior densities of the key parameter pairings under
the monopole+$r(z)$-dipole hypothesis---viz.\ monopole
strength--dipole strength ($m$--$B$) and right ascension--declination
of the dipole vector
($\mathrm{ra}_d$--$\mathrm{dec}_d$)---for error models
$\textsc{\bf{A}}$$\textsc{\bf{A}}$$\textsc{\bf{A}}$,
$\textsc{\bf{A}}_0$$\textsc{\bf{A}}_0$$\textsc{\bf{A}}_0$,
$\textsc{\bf{C}}$$\textsc{\bf{C}}$$\textsc{\bf{C}}$, and
$\textsc{\bf{C}}_0$$\textsc{\bf{C}}_0$$\textsc{\bf{C}}_0$.  While the
posteriors corresponding to each
error model in the ``unbiased'' case
are both highly concentrated around the maximum likelihood
solutions identified by the Webb et al.\ team under their fitting of
the nominal
error model ($\textsc{\bf{A}}_0$$\textsc{\bf{A}}_0$$\textsc{\bf{A}}_0$)---namely,
$\{\mathrm{ra}_d\approx17.5$ hours, $\mathrm{dec}_d\approx-60$
degrees$\}$ and $\{m\approx-0.2\times10^{-5}$,
$B\approx10^{-6}\}$---in the biased (skeptical) case the corresponding
posteriors are evidently more diffuse
and favour markedly a smaller dipole strength.

To discriminate between this plethora of
 plausible generative models with quasi-degenerate parameters---and to
 thereby identify the strongest candidate(s)
 for explaining the apparent spatial variation of the quasar dataset---we
 now proceed
 to the Bayesian comparison of marginal likelihoods (cf.\
 \citealt{kas95}).

\section{Reverse Logistic Regression}\label{rlr}
For each hypothesis--error model pairing we estimate the marginal
likelihood based on our parallel tempered MCMC draws using the technique
of reverse logistic regression (RLR) introduced by \citet{gey94}.  A
common mis-conception in astronomical studies is that
despite having performed extensive parallel tempering to simulate from
the Bayesian posterior it remains necessary to apply a further 
sampling technique for model selection purposes.  In fact, RLR (or
the equivalent Density of States summation; \citealt{cam13}) offers a
reliable solution for this scenario, requiring at minimum a set of
likelihoods drawn from the prior\footnote{Or some suitable alternative reference
density; i.e., with
known normalization and overlapping support.} and from one other
temperature configuration, $\pi(\theta)L(\theta)^\beta$, perhaps the
posterior itself ($\beta=1$).  As in our recent update to the
\texttt{MultiNest} code (INS; \citealt{fer13}) the RLR algorithm uses
a `losing the labels' strategy in which the likelihoods from all
specified proposals are pooled together and a recursive computation used to
 estimate normalizations simultaneously for all except the prior (or
 other reference density) for which $Z_{\beta(1)=0} = 1$ is assumed known. Though we give below only the RLR formula for the case of parallel
tempering between the prior and posterior,  the interested reader may
refer to our recent review paper \citep{cam13} for the general
formula and its derivation, as well as
 a variety of novel implementation suggestions.  

Supposing one has drawn a series
of $\theta_i^{(j)} \sim \pi(\theta)L(\theta)^{\beta(j)}$ for $i =
1,\ldots,n(j)$ and $j=1,\ldots,m$ with $\beta(1) = 0$ and $\beta(m) = 1$, the
RLR estimator for the marginal likelihood of the posterior, $\hat{Z}\
(=\hat{Z}_m)$, and intermediate (bridging) densities, $\hat{Z}_j$ ($1
< j < m$), may be recovered via
recursion over the following update equation:
\begin{eqnarray}
\hat{Z}_j = \sum_{j=1}^m \sum_{i=1}^{n(j)}\left( L(\theta_i^{(j)})^{\beta(j)}/[\sum_{s=1}^m
n(s) L(\theta_i^{(j)})^{\beta(s)}/\hat{Z}_{s}] \right)\\
= \sum_{i=1}^n \left(  L(\theta_i)^{\beta(j)}/[\sum_{s=1}^m
n(s) L(\theta_i)^{\beta(s)}/\hat{Z}_{s}]   \right).
\end{eqnarray}
The second line of this equation corresponds to the observation that
recovery of the marginal likelihood by this method does not in fact
require knowledge of which temperature a particular sample point has been
sampled from; only the sampling design itself (the set of $\beta(j)$
and their corresponding $n(j)$) need be known. This is consistent with the description of RLR as a `losing the labels'
type scheme (see \citealt{kon03} for a detailed discussion).  For
uncertainty estimation \citet{gey94}
gives one expression for the asymptotic variance in the RLR
solution, while \citet{kon03} give another structured for improved
computational efficiency.

Finally, as stressed in \citet{cam13}, it is important to note the
power of parallel tempering with RLR for prior-sensitivity analysis.
Having explored well beyond the confines of the highest posterior density
region at some $\beta(j) < 1$, for which (via RLR) we now have
estimated normalization constants, we are well placed to recompute
$\hat{Z}$ under a range of  alternative prior densities through importance
sample reweighting.  In this respect we treat the pooled draws from parallel
tempering as a (pseudo-)importance sampling proposal density,
\begin{equation}
g(\theta_i) = \sum_{j=1}^m (n[j]/n) [\pi(\theta_i)
L(\theta_i)^{\beta(j)}] / \hat{Z}_{j}.
\end{equation}   
Provided the support of the alternative prior,
$\pi_\mathrm{alt}(\cdot)$, is contained within that of the default
 we have the simple unbiased estimator,
\begin{equation}
\hat{Z}_\mathrm{alt} = \sum_{i=1}^{n} \pi_\mathrm{alt}(\theta_i)
L(\theta_i) / g(\theta_i) / n.
\end{equation}
Computation of which requires no further likelihood evaluations, only
the calculation of $\pi_\mathrm{alt}(\theta_i)$ for each $i=1,\ldots,n$.
The uncertainty of this importance sampling estimator will not be much
greater than that of our original RLR solution provided the
divergence between $\pi_\mathrm{alt}(\cdot)$ and $\pi(\cdot)$ is not
overwhelming---a condition easily monitored via the effective sample
size \citep{kon03}.  We exploit this valuable property of the RLR
estimator to examine the robustness of our BSM results in Section \ref{marginallikelihoods} below.

\section{Marginal Likelihoods}\label{marginallikelihoods}
Marginal likelihoods were computed for ten alternative permutations of
our four candidate error
forms (Section \ref{errormodels}) over the three
$\varepsilon_\mathrm{sys}$ groups of the quasar dataset---in particular, the four general (skeptical) error forms applied
homogeneously
($\textsc{\bf{A}}$$\textsc{\bf{A}}$$\textsc{\bf{A}}$,
$\textsc{\bf{B}}$$\textsc{\bf{B}}$$\textsc{\bf{B}}$,
$\textsc{\bf{C}}$$\textsc{\bf{C}}$$\textsc{\bf{C}}$,
$\textsc{\bf{D}}$$\textsc{\bf{D}}$$\textsc{\bf{D}}$), their
``unbiased'' case counterparts applied likewise ($\textsc{\bf{A}}_0$$\textsc{\bf{A}}_0$$\textsc{\bf{A}}_0$,
$\textsc{\bf{B}}_0$$\textsc{\bf{B}}_0$$\textsc{\bf{B}}_0$,
$\textsc{\bf{C}}_0$$\textsc{\bf{C}}_0$$\textsc{\bf{C}}_0$,
$\textsc{\bf{D}}_0$$\textsc{\bf{D}}_0$$\textsc{\bf{D}}_0$), plus two
non-homogeneous permutations ($\textsc{\bf{C}}$$\textsc{\bf{A}}$$\textsc{\bf{D}}$ and
$\textsc{\bf{C}}$$\textsc{\bf{A}}$$\textsc{\bf{A}}$) motivated by the
observed distributions of residuals $(\Delta \alpha / \alpha-\Delta
\alpha / \alpha{}_\mathrm{mod})$ about the \citet{kin12} dipole fit along the equatorial
strip of Figure \ref{kingerrors}---under each of
the three competing hypotheses---namely, strict null, monopole-only null, and
monopole+$r(z)$-dipole.  These results, totalling thirty separate
marginal likelihood estimates, are compiled (in log $\hat{Z}$ format) in Table
\ref{MLtable}.

\begin{table*}
\renewcommand{\arraystretch}{1.4}

\caption{Log Marginal Likelihoods for Each Error Model--Hypothesis
  Pairing Tested Against the Quasar Dataset}
\label{MLtable}
\begin{tabular}{lccccc}
\hline
 & $\textsc{\bf{AAA}}$ & $\textsc{\bf{A}}_0$$\textsc{\bf{A}}_0$$\textsc{\bf{A}}_0$ & $\textsc{\bf{BBB}}$ &
 $\textsc{\bf{B}}_0$$\textsc{\bf{B}}_0$$\textsc{\bf{B}}_0$ & $\textsc{\bf{CCC}}$\\
\hline
Strict Null & \fcolorbox{Periwinkle}{Goldenrod}{-608.0} & -616.0 & -617.3 &
-629.2 & -608.2\\

Monopole & -608.5 & -615.9 & -618.2 &
-630.5 & -608.7 \\

Dipole & -610.5 & -610.3 &  -619.1 & -623.1 & -608.9\\
\hline
  &$\textsc{\bf{C}}_0$$\textsc{\bf{C}}_0$$\textsc{\bf{C}}_0$ &
 $\textsc{\bf{DDD}}$  &
 $\textsc{\bf{D}}_0$$\textsc{\bf{D}}_0$$\textsc{\bf{D}}_0$ &
 $\textsc{\bf{CAD}}$  &  $\textsc{\bf{CAA}}$\\
\hline
Strict Null & -616.9 & -626.3 & -627.3 & -614.6 & \fcolorbox{Periwinkle}{Goldenrod}{-607.7}\\

Monopole & -614.6 & -626.5 & -629.2 & -615.4 & -608.0\\

Dipole & \fcolorbox{White}{Goldenrod}{-609.4} &
-617.8 & -617.8 & -614.8 & -608.2\\

\end{tabular}
\medskip
\\
All quoted
  log $\hat{Z}$ have uncertainties $\lesssim$ 0.1.
\end{table*}

The first
point we note here upon consideration of the above is that for the
nominal error model ($\textsc{\bf{A}}_0$$\textsc{\bf{A}}_0$$\textsc{\bf{A}}_0$) of the Webb et al.\ team---in which the unexplained
source of error in each $\varepsilon_\mathrm{sys}$ group is assumed
strictly Normal and strictly unbiased (zero mean and mode)---the log marginal likelihood of the dipole
hypothesis ($\log \hat{Z} = -610.3$) exceeds that of both the strict null ($\log
Z = -616.0$) and monopole-only null ($\log \hat{Z} = -615.9$)
alternatives  by $\Delta \log \hat{Z} \approx 5.9$.  Under the Jeffreys scale for interpretation of
Bayes factors ($\mathrm{B.F.}=\exp[\Delta \log \hat{Z}]$)
such an extreme result (a $\mathrm{B.F.} > 100$) should be
considered decisive evidence in favour of the former; broadly
 consistent at face value (i.e., neglecting for now both our results
 for alternative error models and
 any subjective priors one might hold on the relative merits of these hypotheses) with the Webb et al.\ team's bootstrap
randomization-based estimate of $\sim$4$\sigma$ support for 
spatial variation in the quasar dataset.  (We say ``broadly'' as, of
course, one cannot directly compare $p$-values
against Bayes factors without an explicit calibration of the latter.)

A further inspection of the results compiled in Table \ref{MLtable}
reveals that this
particular 
ordering of the null and dipole hypotheses by marginal likelihood recovered
under the nominal error model
($\textsc{\bf{A}}_0$$\textsc{\bf{A}}_0$$\textsc{\bf{A}}_0$) is in fact
conserved across all three
alternative error models tested in their ``unbiased'' cases---namely,
$\textsc{\bf{B}}_0$$\textsc{\bf{B}}_0$$\textsc{\bf{B}}_0$ (heavy
tailed),
$\textsc{\bf{C}}_0$$\textsc{\bf{C}}_0$$\textsc{\bf{C}}_0$ (skew),
and $\textsc{\bf{D}}_0$$\textsc{\bf{D}}_0$$\textsc{\bf{D}}_0$
(bimodal).  Interestingly, one may also note that the greatest
marginal likelihood recovered here for the (\textit{prima facie}) preferred dipole
hypothesis under any of these error models is that of log $\hat{Z} = -609.4$
for the skew Normal form; that is, even with the Ockham's razor-like
tendency of the Bayes factor to favour simpler models (cf.\
\citealt{jef92,jay03}) we observe here a slight posterior preference
for this asymmetric (but zero mode) density relative to its default symmetric  alternative.

Turning now to the results in Table \ref{MLtable} for our general (skeptical) error models, which expressly allow for
the possibility of opposing systematic biases between the observations
from the Keck and VLT telescopes, we discover for all but one
permutation tested ($\textsc{\bf{D}}\textsc{\bf{D}}\textsc{\bf{D}}$) a reversal of the above
ranking of hypotheses; that is, for error models
$\textsc{\bf{A}}\textsc{\bf{A}}\textsc{\bf{A}}$,
$\textsc{\bf{B}}\textsc{\bf{B}}\textsc{\bf{B}}$,
$\textsc{\bf{C}}\textsc{\bf{C}}\textsc{\bf{C}}$,
$\textsc{\bf{C}}\textsc{\bf{A}}\textsc{\bf{D}}$, and
$\textsc{\bf{C}}\textsc{\bf{A}}\textsc{\bf{A}}$ the strict null and monopole-only null 
hypotheses are in fact preferred in marginal
likelihood over the Webb et al.\ team's monopole+$r(z)$-dipole hypothesis---although we note
that the strength of this preference ranges only between ``barely worth
mentioning'' and ``substantial'' on the Jeffreys scale.  Furthermore,
for three of these error models
($\textsc{\bf{A}}\textsc{\bf{A}}\textsc{\bf{A}}$,
$\textsc{\bf{C}}\textsc{\bf{C}}\textsc{\bf{C}}$, and
$\textsc{\bf{C}}\textsc{\bf{A}}\textsc{\bf{A}}$) the resulting
marginal likelihoods for all hypotheses are consistently greater than
the maximum recovered for the dipole hypothesis under any unbiased
error model; with the greatest maximum
likelihood of the present study (log $\hat{Z} = -607.7$) recovered for 
$\textsc{\bf{C}}\textsc{\bf{A}}\textsc{\bf{A}}$ under the strict null
hypothesis.  That we recover such support for a skew Normal term in
the Keck LC $\varepsilon_\mathrm{sys}$ group is quite intriguing in that
it may offer some insight into the origin of the unexplained errors of
this dataset---in particular, it suggests a sole contributor, such as
bias in the Keck wavelength calibration solution, rather than the
accumulation of many disparate sources converging toward a Normal.
Nevertheless, for simplicity, we have elected to focus the ensuing
discussion on the almost-as-successful (log $\hat{Z} = -608.0$) $\textsc{\bf{AAA}}$
model as this was our \textit{a priori}
default skeptical proposal and is most similar in form to its (unbiased)
nominal rival.

\begin{figure*}
\vspace{0pc}
\includegraphics[width=0.825\textwidth]{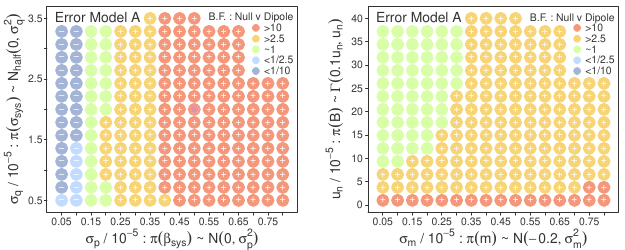}
\vspace{-0.0cm}\caption{Exploring the prior-sensitivity of the Bayes
  factor (B.F.) comparing the strict null against the 
  monopole+$r(z)$-dipole hypothesis under the
   skeptical error model, $\textsc{\bf{AAA}}$. By importance
  sample re-weighting of our tempered likelihood draws (the
  normalizations of which were estimated during our earlier RLR run)
  we can rapidly recompute the log $\hat{Z}$ for each hypothesis, and hence the
  relative Bayes factor, under alternative
  priors, $\pi_\mathrm{alt}(\cdot)$, with supports enclosed by those of the original.  In the lefthand panel we
  examine the impact of varying our hyperparameters, $\sigma_p$ and $\sigma_q$, controlling the 
    expected absolute size of the  bias 
    and the standard deviation ($\beta_\mathrm{sys}$ and $\sigma_\mathrm{sys}$, respectively) in the
    unexplained error term, $\varepsilon_\mathrm{sys}$.  For
    reference, our original prior corresponds to $\sigma_p = 0.5\times
    10^{-5}$ and $\sigma_q = 2.0\times
    10^{-5}$. In the
    righthand panel we apply alternative priors 
    concentrated to varying degrees on the (\textit{a posteriori}) `best-fit' values of the
    monopole and dipole strength terms ($m$ and $B$, respectively) in 
    the dipole hypothesis.  In each case it is
  evident that large modifications of our stated prior beliefs
  are required
  to overturn the weight of evidence against spatial variation in
  $\Delta \alpha / \alpha$ under this biased error model.}
\label{prioreffect}
\end{figure*}

As mentioned earlier, in order to investigate the sensitivity of these results to our choice of
priors on both error model and hypothesis parameters we make use here of 
 an important property of the RLR algorithm---namely, that it allows for rapid recomputation  of the
 marginal likelihood  (requiring no new
 likelihood evaluations) under mild modifications to the prior
 density.  In Figure
 \ref{prioreffect} we qualify the subjectivity in this
 regard of our recovered Bayes
  factors (B.F.) comparing the strict null against the
  monopole+$r(z)$-dipole hypothesis under the biased error model,
  $\textsc{\bf{A}}\textsc{\bf{A}}\textsc{\bf{A}}$.  In the lefthand panel we
  examine the impact of varying our hyperparameters, $\sigma_p$ and
  $\sigma_q$ (nominally, $\sigma_p = 0.5\times
    10^{-5}$ and $\sigma_q = 2.0\times
    10^{-5}$), controlling the 
    expected absolute size of the bias, $\beta_\mathrm{sys}$, 
    and the standard deviation, $\sigma_\mathrm{sys}$, respectively, for the
    unexplained error term.  In the
    righthand panel we apply alternative priors 
    concentrated to varying degrees on the (\textit{a posteriori}) `best-fit' values of the
    monopole and dipole strength terms, $m$ and $B$ respectively, of
    the dipole hypothesis.  In each case it is
  evident that large modifications of our stated prior beliefs are required
  to overturn the weight of evidence against spatial variation in the
  fine structure constant under this biased error model.  Either we must drastically 
  rescale our prior density on the magnitude of the possible
  error bias to strongly favour biases below the $\sim$10$^{-6}$
  level, or else provide a convincing post-hoc theoretical
  justification for the precise strengths of the dipole model's  
 $m$ and $B$ terms.  

\section{Conclusions}\label{conclusions}
Although there exist a number of well-developed 
cosmological theories within which a time- and/or space-varying fine structure constant can be readily admitted
\citep{bek82,car98,mar84,bra03}, for many physicists the prior
probability of such must be considered small given both our faith in certain long-standing
physical principles and the null results of 
previous experiments designed to test these.  In particular, past analyses of the Oklo
natural fission reactor \citep{shl76,dam96,gou06}, Earth-fallen meteorite samples
\citep{oli04}, and optical atomic clocks \citep{for07,ros08} have strongly
favoured a scenario of negligible \textit{temporal} variation in
$\alpha$ locally; with
\citet{ros08}, for instance, determining $|\Delta \alpha/\alpha|<1.6 \times 10^{-17}$ per
year at present on Earth.  These results may ``daringly'' (with
respect to the strong assumption of zero higher derivatives) be
extrapolated to $|\Delta
\alpha/\alpha| < 2\times10^{-7}$ since the Big Bang ($\sim$13.4 Gyr ago); with
more prosaic but reliable constraints of $|\Delta \alpha / \alpha| < 0.1$
over this time provided by contemporary analyses of fluctuations in the Cosmic
Microwave Background \citep{roc04,nak08,men09,gal10,cal11}.

\textit{Spatial} variation
in the fine structure constant, on the other hand, has been previously
constrained to $|\Delta \alpha/\alpha| < 10^{-4}$ on cosmic scales via both
emission and absorption line based quasar studies
\citep{bah65,bah67,iva99}. Although the Webb et al.\ team's claimed
dipole strength contributes an effect well below this
level the earlier absorption line results may nevertheless be supposed
to have reinforced the
prior beliefs of many astronomers (by way of the cosmological
principle) that
the physical laws of the Universe are spatially invariant.  A related
expectation that the Universe should appear essentially homogeneous when
viewed on sufficiently large scales has also been hitherto
well-supported by measurements of the fractal dimension (converging
toward three) at large
scales in galaxy redshift surveys \citep{mar98,scr12} and by the
isotropy of the Cosmic Microwave Background (\citealt{man86}; though
see \citealt{eri07} and \citealt{cla99}).  The proposal of a
large-scale spatial variation of the fine structure constant across the
observable Universe runs into direct conflict with this
established paradigm.

To quantify our inherent preference for the strict null hypothesis over the dipole
 we might therefore, for argument's sake, suppose a prior
log odds
ratio of 5, such that $P_\mathrm{prior}(\mathrm{strict\
  null})/P_\mathrm{prior}(\mathrm{dipole}) \approx 150$.  In this conservative scenario,
even the Bayes factor of $\sim$$300$ in favour of the dipole under a
strictly unbiased, Normal error model is reduced to a posterior odds
ratio of just $2:1$, providing insufficient evidence to convince
ourselves beyond doubt that the fine structure 
constant does indeed vary across the Universe in the manner described.
Such a prior hypothesis weighting against the dipole also appears reflected
in the relative caution with which both independent cosmologists \citep{oli11} and the
Webb et al.\ team themselves have at times discussed their results.
Moreover, as we have demonstrated herein (cf.\ Section \ref{marginallikelihoods}) the skeptical interpretation of null variation under a \textit{biased} error
model (in which the current $\Delta \alpha / \alpha$ estimates from
the Keck and VLT carry systematic biases of opposing sign) may in fact
be preferred from a Bayesian model selection perspective.
Thus new
observations are undoubtably required if the Webb et al.\ team or
others are to convince a majority of cosmologists to accept the dipole hypothesis.

   To this end a dedicated campaign targetting quasars \textit{at the
     poles of the inferred dipole} has 
been proposed \citep{web11,kin12}, and is, in effect, already in progress  with
a first series of high resolution spectra suitable for $\Delta
\alpha/\alpha$ estimation recently obtained on the VLT under Large
Program 185.A-0745
\citep{mol13}.  Assuming these measurements are 
subject to both explained and unexplained error terms consistent with those
of the original quasar dataset---and supposing further the availability of an
equivalent new Keck sample---one can easily forecast the power of
such an experimental strategy (and potential alternatives) for settling the null vs.\ dipole
debate.  One way to do this is via Monte Carlo simulation of 
future Bayes factors, assuming the dipole plus (unbiased) Normal error model pairing as
truth.  

We perform such computational simulations here by repetition over the
following procedure.   First, we 
draw a $\{\bm{\theta}_m,\bm{\theta}_e\}$ pair of dipole hypothesis
plus $\textsc{\bf{A}}_0$$\textsc{\bf{A}}_0$$\textsc{\bf{A}}_0$ error
model parameter vectors from the current posterior.  For each
of $n_\mathrm{new}$ proposed targets we draw a  
$\Delta \alpha/\alpha_\mathrm{obs}$ estimate accordingly with
reference to its nominated
telescope (i.e., $\epsilon_\mathrm{sys}$ group), sightline, and redshift, plus a $\sigma_\mathrm{obs}$ sampled (with replacement)
from the current quasar dataset.
We then
compute the marginal log likelihoods of these mock observations under
both 
the strict null and dipole hypotheses, proposing a biased error model for
the former and an unbiased
error model for the latter, and treating our
current posteriors for each as priors.  In the case of limited
new data  this computation can be efficiently and accurately performed  via
importance sampling from our earlier reverse logistic regression
draws;  that is, for $n_\mathrm{new}$ small enough the
  updated posterior will not differ too markedly from the current
  posterior.  For reference we also perform these simulations with
  unbiased error models assigned to both hypotheses.  The distribution of future 
Bayes factors under a proposed observing strategy  approximated in
this manner 
can be termed the ``predicted posterior odds distribution'' (or PPOD; \citealt{tro07}).

\begin{figure*}
\vspace{0pc}
\includegraphics[width=0.825\textwidth]{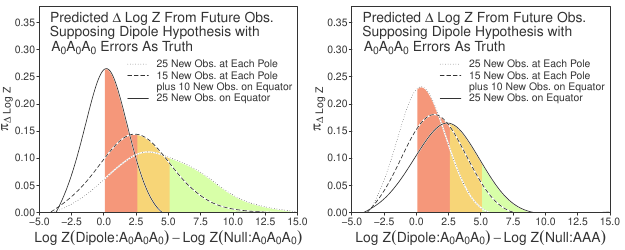}
\vspace{-0.05cm}\caption{The predicted power of future quasar 
  observations for increasing the log Bayes factor in favour of the
  dipole hypothesis (supposed here as truth) under the nominal (unbiased, Normal) error model
  in the lefthand panel and under our default skeptical (biased,
  Normal) error model in the righthand panel.  In each case we suppose
  the availability of 25 new $\Delta\alpha/\alpha$ estimates from each
  of the VLT and Keck telescopes subject to both explained and
  unexplained error terms consistent with those of the original quasar
  dataset.  Three alternative targetting strategies have been
  simulated: \textsc{(i)} all targets  placed at the
current maximum a posteriori sightline of the pole (for the VLT)
or anti-pole (for the Keck); \textsc{(ii)} all targets placed on the dipole's
equator; and \textsc{(iii)} a compromise
 with ten measurements on the equator and the remainder
on the pole or anti-pole (as appropriate).  Note that the future log
posterior odds ratio for each case should be considered the sum of the future log Bayes factor
 considered above with the current log Bayes factor
 from the original quasar dataset (lefthand panel: $+$5.7; righthand
 panel: $-$2.5) minus our log prior odds ratio of 5
 against the dipole.}
\label{future}
\end{figure*}

In Figure
\ref{future} we present the (log) PPOD for three basic targetting
strategies, all supposing 25 detected absorbers with $\Delta \alpha /
\alpha$ measurements from observations on each of the Keck and VLT
telescopes.\footnote{The choice of 25 new absorbers from each
  telescope (i.e., $n_\mathrm{new} = 50$) is consistent with
  expectations for the ongoing VLT Large Program \citep{mol13}.}  In the first, all targets are supposed placed at the
current maximum a posteriori sightline of the pole (for the VLT)
or anti-pole (for the Keck); in the second, all targets are supposed placed on the dipole's
equator (accessible to both telescopes); and in the third a compromise
is made with ten measurements on the equator and the remainder
on the pole or anti-pole (as appropriate).  As expected it is indeed
the first strategy that gives the greatest chance of recovering the
desired additional support of a further $+$5 on the log Bayes factor
of the null vs.\ dipole comparison with the unbiased error model
applied to both.  The mixed targetting strategy performs a little
worse than the dipole-only case, while the equator-only strategy
performs worst of all.  With regards to choosing between the null and dipole when allowing for
a biased error model on the former the performance of our three
strategies is in fact reversed, such that the equatorial design offers the greatest chance of a
conclusive outcome.  An intuitive interpretation of this result is simply that for
distinguishing between a biased and unbiased error model under two
competing hypotheses the most effective design is to observe at locations where those underlying hypotheses are already in
close agreement.  

Finally it is worth noting the wide range of predicted Bayes factors
under all these possible observating strategies, which are so broad as
to include a non-trivial possibility of rejecting the true
hypothesis despite an optimal experimental design.  
This phenomenon arises principally from the broad range allowed for
the strength parameter of the dipole, $B$, under the current posterior
(see Figure \ref{mcmcMB}), which gives non-negligible weight to the
possibility that the dipole signal is far too small (with respect to
the observational errors) to confidently recover from an $n_\mathrm{new}=50$ sample.

Thus, we reach the final conclusions of our Bayesian reanalysis of the
quasar dataset.  Namely, that: \textsc{(i)} given both our incomplete understanding of the
observational errors and our limited
theoretical (prior) expectations regarding the properties of any spatial
variation in the fine structure constant, the present observational coverage (featuring
limited overlap of the two telescopes for which opposing biases might
be suspected) must be deemed inadequate to properly distinguish the
Webb et al.\ team's proposed dipole field from the (strict or
monopole) null; and \textsc{(ii)} one cannot afford to
overlook the importance of observations along
the equator of the alleged dipole in additon to those proposed along
its poles when planning future campaigns with the Keck and VLT telescopes for the purpose of settling this debate.

We have also demonstrated in this study the utility of the reverse
logistic regression technique for marginal likelihood computation
with efficient prior-sensitivity analysis.

\section*{Acknowledgments}
\noindent\texttt{[1]} E.C.\ is grateful for financial support from the Australian Research Council.

\label{lastpage}
\end{document}